\documentclass[twocolumn]{aastex7}
\usepackage{xcolor}
\usepackage{caption}
\usepackage{subcaption}
\usepackage{graphicx}
\usepackage{array,multirow,graphicx}
 \usepackage{float}
 \usepackage{supertabular}
 \usepackage{amssymb}

\usepackage[T1]{fontenc}

\DeclareRobustCommand{\VAN}[3]{#2}
\let\VANthebibliography\thebibliography
\def\thebibliography{\DeclareRobustCommand{\VAN}[3]{##3}\VANthebibliography}

\usepackage{longtable}
\usepackage{graphicx}	
\def\code#1{\texttt{#1}}

\usepackage{newtxtext,newtxmath}

\begin{document}

\title{Optical Spectroscopy Reveals Hidden Neutron-capture Elemental Abundance Differences among \textit{APOGEE}-identified Chemical Doppelg{\"a}ngers\footnote{ This paper includes data taken with the Harlan J. Smith 2.7m telescope at The McDonald Observatory of The University of Texas at Austin.}}

\author[orcid=0000-0002-0900-6076,sname='Manea']{Catherine Manea}
\altaffiliation{NSF Astronomy and Astrophysics Postdoctoral Fellow}
\affiliation{Columbia Astrophysics Laboratory, Columbia University, New York, NY, 10027, USA}
\email[]{cm4582@columbia.edu}  

\author[orcid=0000-0001-5082-6693]{Melissa Ness}
\affiliation{Research School of Astronomy and Astrophysics, Australian National University, Canberra, ACT 2601, Australia} 
\email[]{x}

\author[orcid=0000-0002-1423-2174]{Keith Hawkins}
\affiliation{Department of Astronomy, The University of Texas at Austin, Austin, TX 78712, USA}
\email[]{x}

\author[orcid=0000-0003-1124-8477]{Greg Zeimann}
\affiliation{Hobby–Eberly Telescope, University of Texas at Austin, Austin, TX 78712, USA} 
\affiliation{McDonald Observatory, The University of Texas at Austin, Austin, TX 78712, USA}
\email[]{x}

\author[orcid=0000-0003-2866-9403]{David W. Hogg}
\affiliation{Center for Cosmology and Particle Physics, Department of Physics, New York University, 726 Broadway, New York, NY 10003, USA}
\affiliation{Max-Planck-Institut für Astronomie, Königstuhl 17, D-69117 Heidelberg, Germany}
\affiliation{Center for Computational Astrophysics, Flatiron Institute, 162 Fifth Ave., New York, NY 10010, USA} 
\email[]{x}

\author[orcid=0000-0001-5522-5029]{Carrie Filion}
\affiliation{Center for Computational Astrophysics, Flatiron Institute, 162 Fifth Ave., New York, NY 10010, USA} 
\email[]{x}

\author[orcid=0000-0001-9345-9977]{Emily J. Griffith}
\altaffiliation{NSF Astronomy and Astrophysics Postdoctoral Fellow}
\affiliation{Center for Astrophysics and Space Astronomy, Department of Astrophysical and Planetary Sciences, University  of Colorado, 389~UCB, Boulder,~CO 80309-0389, USA}
\email[]{x}

\author[orcid=0000-0001-6244-6727]{Kathryn Johnston}
\affiliation{Columbia Astrophysics Laboratory, Columbia University, New York, NY, 10027, USA} 
\email[]{x}

\author[orcid=0000-0003-0174-0564]{Andrew Casey}
\affiliation{Center for Computational Astrophysics, Flatiron Institute, 162 Fifth Ave., New York, NY 10010, USA}
\affiliation{School of Physics and Astronomy, Monash University, Melbourne, VIC 3800, Australia}
\affiliation{ARC Centre of Excellence for All Sky Astrophysics in Three Dimensions (ASTRO-3D), Australia}
\email[]{x}

\author[orcid=0000-0002-3855-3060]{Zoe Hackshaw}
\affiliation{Department of Astronomy, The University of Texas at Austin, Austin, TX 78712, USA}
\email[]{x}

\author[orcid=0000-0003-3707-5746]{Tyler Nelson}
\affiliation{Department of Physics, University of Southern Maine, Portland, ME 04104}  
\email[]{x}

\author[sname='Marks']{Micah Marks}
\affiliation{Department of Astronomy, The University of Texas at Austin, Austin, TX 78712, USA}
\email[]{x}

\begin{abstract} 
Grouping stars by chemical similarity has the potential to reveal the Milky Way's evolutionary history.  The \textsl{APOGEE} stellar spectroscopic survey has the resolution and sensitivity for this task.  However, \textsl{APOGEE} lacks access to strong lines of neutron-capture elements (Z\textgreater28) which have nucleosynthetic origins that are distinct from those of the lighter elements.  We assess whether \textsl{APOGEE} abundances are sufficient for selecting chemically similar disk stars by identifying 25 pairs of chemical ``doppelg{\"a}ngers" in \textsl{APOGEE} DR17 and following them up with the \textsl{Tull} spectrograph, an optical, R$\sim$60,000 echelle on the McDonald Observatory 2.7-m telescope.  Line-by-line differential analyses of pairs' optical spectra reveals neutron-capture (Y, Zr, Ba, La, Ce, Nd, and Eu) elemental abundance differences of $\Delta$[X/Fe] $\sim$ 0.020$\pm$0.015 to 0.380$\pm$0.15 dex (4-140\%), and up to 0.05 dex (12\%) on average, a factor of 1-2 times higher than intra-cluster pairs. This is despite the pairs sharing nearly identical \textsl{APOGEE}-reported abundances and [C/N] ratios, a tracer of giant-star age. This work illustrates that even when \textsl{APOGEE} abundances derived from SNR$>300$ spectra are available, optically-measured neutron-capture element abundances contain critical information about composition similarity.  These results hold implications for the chemical dimensionality of the disk, mixing within the interstellar medium, and chemical tagging with the neutron-capture elements.
\end{abstract}

\keywords{\textsl{s}-process (1419), Surveys (1671), Chemical abundances (224), Galaxy chemical evolution (580)}


\section{Introduction}\label{sec:intro}
The chemical composition of some volume of the Galactic interstellar medium (ISM) is set by the legacy of chemical enrichment events (e.g., stellar nucleosynthesis and feedback, mixing, accretion) that came prior.  Stars form from the ISM and, in the absence of internal evolutionary processes such as dredge up and atomic diffusion, their surface compositions generally reflect the composition of the ISM from which they formed.  Even as the surrounding ISM continues to evolve, stars retain a chemical memory of their birth environment, making them powerful tools for studying the Milky Way's fundamental processes.  Spectroscopic surveys, such as Apache Point Observatory Galactic Evolution Experiment (\textsl{APOGEE}, e.g., \citealp{APOGEEDR17}), GALactic Archaeology with HERMES (\textsl{GALAH}, e.g., \citealp{Buder2018, GALAHDR3}), \textit{Gaia}-ESO (e.g.,\citealp{Gilmore2022, Randich2022}), and Large Sky Area Multi-Object Fiber Spectroscopic Telescope Medium Resolution Survey (\textsl{LAMOST MRS}, e.g., \citealp{LAMOSTMRS}), are collecting a combined millions of medium-resolution (R$\sim$7,500 to R$\sim$28,000 depending on the survey) stellar spectra that contain chemical information for up to 30 elements.  With such a large quantity of chemical data, we are well-poised to explore chemical trends in the Galaxy in order to probe its star formation history, past accretion events, mixing mechanisms, and stellar nucleosynthesis.

Grouping stars by chemical similarity is fundamental to many Galactic studies that leverage stellar abundances to disentangle our Milky Way's evolutionary history and present day structure.  For example, the seminal works of e.g., \citealt{Roman1952, Wallerstein1962, Tinsley1968, Tinsley1979} led to the discovery of a chemical bimodality in the Galactic disk where stars can be grouped into high and low $\alpha$ (e.g., O, Mg, Ca, and Si) populations \citep[e.g.,][]{Gratton1996, Fuhrmann1998, Reddy2003, Bensby2011, Bovy2012, Anders2014, hayden15, Buck2020, Imig2023}.  Stellar abundances also aided in the discovery of the \textit{Gaia}-Sausage-Enceladus system, a population of accreted stars in the inner Galactic halo that show distinct compositions (low Fe and low $\alpha$ among other chemical trends) and is understood to be the last major merger experienced by our Galaxy \citep[e.g.,][]{belokurov18, helmi18, Haywood2018, Hayes2018, Vincenzo2019, Carrillo2022, Carrillo2024, Ciuca2024}.  The limit to which we can meaningfully group stars using chemistry alone has been a major question in the field in the last two decades.  Understanding these limits in the disk is particularly important as it harbors the outcome of \textit{in-situ} star formation.  One potential limiting case is the separation of disk stars into their natal birth groups using chemistry alone \citep[e.g.,][]{Freeman2002, Bland-Hawthorn2010}, a method called \textit{strong chemical tagging}. Even in the limit of negligible uncertainties, strong chemical tagging relies on (i) individual birth sites having unique mean chemical abundance vectors and (ii) intra-cluster abundance scatter that is small enough to be sufficiently discriminatory between clusters.  While the literature agrees that stars in open clusters are generally highly chemically homogeneous \citep[e.g.,][with some exceptions due to stellar evolution effects, e.g. \citealp{Liu2016, Souto2018, Souto2019}]{Bovy2016b, Ness2018, Poovelil2020, Kos2021, Sinha2024}, open clusters at fixed age and birth radius tend to overlap in chemical composition \citep[e.g.,][]{Casamiquela2021, Spina2022, Sinha2024}.  This is supported by, e.g., \citealt{Ness2018, deMijolla2021, Manea2024}, which find that the so-called \textit{chemical doppelg{\"a}nger rate}, the rate at which random pairs of field disk stars are as chemically similar as pairs of intra-cluster stars, is too high for strong chemical tagging to be feasible.  As such, recent efforts have scaled back and instead focused on chemically grouping disk stars by similarity in Galactic birth radius and age or dynamical behavior.  For example, \citet{Ness2019} finds that [Fe/H] abundance and stellar age are enough to predict within 0.02 dex the chemical profile and orbit of an \textsl{APOGEE} DR14 \citep{APOGEEDR14} low $\alpha$-disk star.  This result suggests that [Fe/H] and age alone are sufficient for finding stars with shared chemodynamical behaviors and thus Galactic origins in the low alpha disk.  Numerous other works group stars by chemical similarity to study the Galactic disk \citep[e.g.,][]{Sun2020, PJ2020, Cheng2021, Price-Whelan2021, ReFi2021, Feuillet2022, Lian2022, Ortigoza2023, Foster2024, Horta2024}.

Separating stars into chemically similar populations is evidently an important tool for Galactic science.  However, more commonly than not, the elements considered in such studies are restricted to the light (e.g., C, N), $\alpha$ (e.g., O, Mg, Ca, Si), odd-Z (e.g., Na, Al, Mn), and iron-peak (e.g., Fe, Ni, Cr) elements.  Elements primarily produced by the neutron-capture process (i.e., atomic number greater than 28) are paid comparatively less attention in this context because their lines are often weak and blended (especially in metal rich stars) and thus difficult to measure.  Additionally, most neutron-capture lines lie at blue wavelengths ($\lambda$ \textless 5500 $\AA$), a wavelength regime at which many detectors are insensitive and that some spectroscopic surveys do not access.  \textsl{APOGEE}, for example, is an infrared (H band) survey that is often used for large scale Galactic disk studies due to its broad sampling of the disk and generally high abundance precision.  However, \textsl{APOGEE}'s wavelength coverage limits the survey to few, weak neutron-capture lines \citep[e.g., Ce, Nd, Rb, and Yb,][]{Hasselquist2016, Cunha2017} that are typically measured imprecisely, though the BACCHUS Analysis of Weak Lines in \textsl{APOGEE} Spectra Survey (\textsl{BAWLAS}, \citealp{Hayes2022}) has improved our understanding of neutron-capture lines in \textsl{APOGEE} through empirical upper limits, among other things.  By omitting neutron-capture elements from abundance-based studies of the disk, we may be missing information that is only captured by these elements.

It has been established that the neutron-capture elements behave differently in the disk relative to the lighter elements due to their unique production and dispersal timescales and mechanisms \citep[e.g.,][]{Griffith2022, Griffith2024}.  Asymptotic giant branch (AGB) stars produce half of the trans-iron (Z \textgreater 28) elements via the slow neutron-capture process (\textit{s}-process) and release them into the ISM via stellar winds \citep[e.g.,][]{Karakas2014}.  Stellar \textit{s}-process abundance ratios show a strong correlation with stellar age \citep[e.g.,][and references throughout this manuscript]{Simmerer2004, Nissen2015, Carbajo2024, Vitali2024}.  Additionally, some simulations suggest that \textit{s}-process elements produced in AGB stars are dispersed and mixed across smaller volumes in the ISM relative to the remaining elements which are produced in explosive sources \citep[e.g.,][]{Feng2014, Armillotta2018, Krumholz2018, Emerick2020}.  Their localized dispersal could therefore lead their abundances to vary more significantly between stars born within some fixed volume and time.  Recent simulations by \citet{Zhang2025}, however, suggest the opposite, highlighting that the mixing of neutron-capture elements within the ISM is still an open question.  Extreme explosive events such as neutron-star mergers and hypernovae produce the remaining trans-iron elements via the rapid neutron-capture process \citep[\textit{r}-process, e.g.,][]{Kobayashi2020}, and these elements may also show greater star-to-star variations across the disk as they are synthesized in events that are rare but highly productive \citep[e.g.,][]{Cote2017, Recio2021}. Owing to their unique production sites and dispersal mechanisms, \textit{s}- and \textit{r}-process elements might provide additional tracer information about a star's Galactic origin beyond what is captured by the lighter (Z \textless 28) elements.  By omitting these elements, studies may miss additional chemical information that would distinguish otherwise chemically similar stars.  It is critical that we understand whether these elements are important when grouping stars by chemical similarity.  Doing so also informs models of Galactic chemical evolution, neutron-capture nucleosynthesis, and mixing within the ISM.

In this work, we observe and analyze the optical spectra of a set of carefully selected stars to examine the behavior of neutron-capture elements in otherwise chemically similar stars.  We identify chemically similar stars among a high SNR (SNR \textgreater 300) subset of \textsl{APOGEE} DR17 and assess their neutron-capture element similarity via high resolution (R$\sim$60,000), high signal-to-noise (SNR \textgreater 100) optical spectroscopic follow-up.  We perform a high precision, line-by-line differential analysis (see Section \ref{sec:methods}) to determine compositions in the lighter elements accessible by \textsl{APOGEE} as well as Cu, Zn, and seven neutron-capture elements (Y, Zr, Ba, La, Ce, Nd, and Eu) inaccessible to or imprecisely measured by \textsl{APOGEE}.  We then address the following questions:
\begin{enumerate}
    \item Are \textsl{APOGEE}-identified chemically similar stars also similar in the neutron-capture elements?
    \item Why do some, if any, \textsl{APOGEE}-identified chemically similar stars differ in the neutron-capture elements despite possessing otherwise highly similar lighter (Z\textless29) elemental abundances?
    \item Does similarity in \textsl{APOGEE}-determined [Ce/Fe] correlate to similarity in optically-measured [Ce/Fe]?
\end{enumerate}
In addressing these questions, we explore whether \textsl{APOGEE}-measured elements are sufficient for finding chemically similar stars \citep[e.g.,][]{Ness2019} or if neutron-capture elements may provide additional information to better identify stars with similar Galactic origins.  This work informs future studies that rely on grouping stars by chemical similarity.  Furthermore, our results provide an empirical constraint on the spatial and temporal complexity with which neutron-capture elements are synthesized, dispersed, and mixed across the ISM.

This manuscript is organized as follows.  In Section \ref{sec:data}, we describe the acquisition and reduction of high resolution optical spectra of \textsl{APOGEE} DR17-identified chemically similar stars.  In Section \ref{sec:methods}, we describe our method for determining the chemical similarity of these stars via high precision, line-by-line differential analysis performed on the optical spectra.  In Section \ref{sec:results}, we present our results reporting the chemical similarity of the \textsl{APOGEE}-identified chemically similar pairs in both lighter (Z\textless29) and heavier elements.  In Section \ref{sec:discussion}, we place our results in the context of past observational and theoretical works. Additionally, we compare our abundances to those reported by \textsl{APOGEE} and \textsl{BAWLAS}.  We conclude with a summary in Section \ref{sec:conclusion}.

\section{Data}\label{sec:data}
\subsection{Selection of \textsl{APOGEE} Chemical Doppelg{\"a}ngers for Optical Follow Up}\label{sec:apogee_selection}
In this work, we ask whether \textsl{APOGEE}-identified chemically similar stars are also similar when studied in the optical.  Thus, we must first identify chemically similar stars in the \textsl{APOGEE} survey.  The definition of chemical similarity is ambiguous, particularly when considering the ensemble of elemental abundances available in \textsl{APOGEE}, each of which are measured at different precisions.  We adopt the definition of \citet[][hereafter N18]{Ness2018}, which defines chemically similar stars to be pairs of stars that are just as chemically similar as those born together in the same molecular cloud. Field stars that match this description but show no kinematic indication of having formed together are called ``chemical doppelg{\"a}ngers."  N18's primary goal is to measure the chemical doppelg{\"a}nger rate in \textsl{APOGEE}.  To do this, they consider 20 elemental abundance ratios ([Fe/H], [C/Fe], [N/Fe], [O/Fe], [Na/Fe], [Mg/Fe], [Al/Fe], [Si/Fe], [S/Fe], [K/Fe], [Ca/Fe], [Ti/Fe], [V/Fe], [Mn/Fe], [Ni/Fe], [P/Fe], [Cr/Fe], [Co/Fe], [Cu/Fe], and [Rb/Fe]) homogeneously derived from \textsl{APOGEE} DR13 spectra using \textit{The Cannon} \citep{Ness2015}.  They draw random pairs of stars from the field unassociated with known open clusters and moving groups and quantify their abundance similarity using a $\chi^2$ value defined as:
\begin{equation}\label{eq:chisq}
\chi^2_{nn'} = \sum_{i=1}^I \frac{[x_{ni} - x_{n'i}]^2}{\sigma^2_{ni} + \sigma^2_{n'i}}
\end{equation}
where the two stars in a pair are indexed as $n$, $n'$ and $x$, $\sigma$ are their derived abundance and abundance uncertainty in element $i$.  This leads to a global chemical similarity metric for each stellar pair that considers all 20 elements.  Chemical doppelg{\"a}ngers are defined as stellar pairs with $\chi^2$ values less than the median $\chi^2$ value returned by random pairs of stars drawn from the same open cluster.

We generally follow the N18 method to find chemical doppelg{\"a}ngers for follow up, though we deviate slightly from their approach.  Firstly, we work with \textsl{APOGEE} DR17 \citep{APOGEEDR17} data whereas N18 uses DR13 data.  Unlike N18, we do not re-derive \textsl{APOGEE} DR17 abundances using \textit{The Cannon} \citep{Ness2015} and instead work directly with the publicly released DR17 ASPCAP abundances.  Our quality cuts are stricter to ensure that we only work with high-fidelity abundances: we only consider stars with \code{SNR} > 300, 3650 \textless \code{TEFF} \textless 5760, 0.45 \textless \code{LOGG} \textless 3.95, \code{FE\_H\_FLAG} = 0, \code{X\_FE\_FLAG} = 0, and \code{RUWE} \textless 1.4 (to avoid stars in binaries, e.g., \citealp{Belokurov2020}).  When drawing stellar pairs, we require the stars to share $\rm T_{eff}$ within 50 K and log g within 0.1 dex.  We consider the same elements as N18 with the exception of [P/Fe], [Cu/Fe], and [Rb/Fe] because few DR17 stars have unflagged abundances in these elements, and including these elements makes our sample size prohibitively small.  However, unlike N18, we have access to \textsl{APOGEE} DR17-reported Ce, an \textit{s}-process element, so we include it in our selection to test the efficacy of \textsl{APOGEE}'s Ce at identifying stars with similar \textit{s}-process elemental abundances.  We highlight that we also include C and N in our selection of doppelg{\"a}ngers.  Because C/N can trace stellar age in red giants \citep[e.g.,][]{Casali2019, Spoo2022}, including these elements in our selection increases the likelihood that doppelg{\"a}ngers are also similar in age.

As in N18, when searching for chemical doppelg{\"a}ngers, we ensure that we are selecting from the field as opposed to known clusters.  To separate \textsl{APOGEE} DR17 into field and cluster stars, we use the open cluster membership catalog of \citealt[][hereafter CG18]{Cantat-Gaudin2018}.  We consider field stars to be stars that do not exist in the CG18 catalog, and we consider open cluster stars to be stars in the CG18 catalog with a cluster membership probability exceeding 99\%.  This leaves us with 164 open cluster stars and 64,650 field stars.  We then draw three million random pairs of stars from our high-fidelity field sample and compute $\chi^2$ values for each pair using Equation \ref{eq:chisq}.  This process is repeated for all possible intra-cluster (within the same cluster) pairs from open clusters M~67 and NGC 6819 (as in N18) to serve as our reference sample representing the chemical homogeneity of stars born together.  Finally, as in N18, we draw and compute $\chi^2$ values for all inter-cluster (across all clusters) pairs among the CG18 clusters.  As in N18, we define doppelg{\"a}ngers to be field pairs with $\chi^2$ less than or equal to the median $\chi^2$ value for intra-cluster pairs.  This analysis produces a pool of \textsl{APOGEE} chemical doppelg{\"a}ngers from which we can select targets.  Final targets for optical follow-up were selected semi-randomly from this pool, prioritizing bright (V \textless 11) stars to increase data acquisition efficiency.

In addition to following up \textsl{APOGEE}-identified chemical doppelg{\"a}ngers, we also target 11 giant stars in M~67, an open cluster that has been found to be chemically homogeneous within abundance uncertainties when studying its giant stars \citep[e.g.,][]{Bovy2016a, Souto2018, Poovelil2020, Sinha2024}.  These cluster stars will serve as a reference for the chemical homogeneity of conatal stars.  We select stars from M~67 with CG18-assigned membership probabilities \textgreater 99\% that have \textsl{APOGEE} DR17 spectra with SNR \textgreater 100\footnote{Note the lower SNR limit here.  This is because there are insufficient M~67 giants with SNR \textgreater 300 APOGEE spectra.}.  We are not able to observe NGC 6819, the second cluster used in N18, because its giant stars exceed the magnitude limit of the \textsl{Tull} spectrograph on the 2.7m telescope.



\begin{figure}
    \centering
    \includegraphics[width=.5\textwidth]{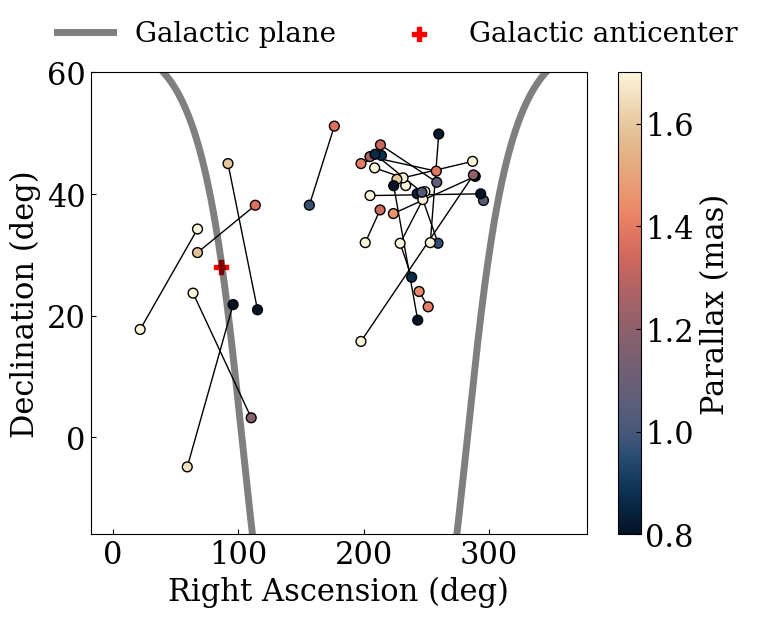}
    \caption{The right ascension vs. declination distribution of our observed sample of \textsl{APOGEE}-identified doppelg{\"a}ngers.  Doppelg{\"a}ngers are connected with a line and colored by their \textit{Gaia} DR3 parallax.  The Galactic plane is marked in gray and the Galactic anticenter is marked by a red plus.  Doppelg{\"a}ngers span a range of on-sky positions, parallaxes, and on-sky separations and do not appear to be kinematically related.}
    \label{fig:radec}
\end{figure}

\begin{figure}
    \centering
    \includegraphics[width=.5\textwidth]{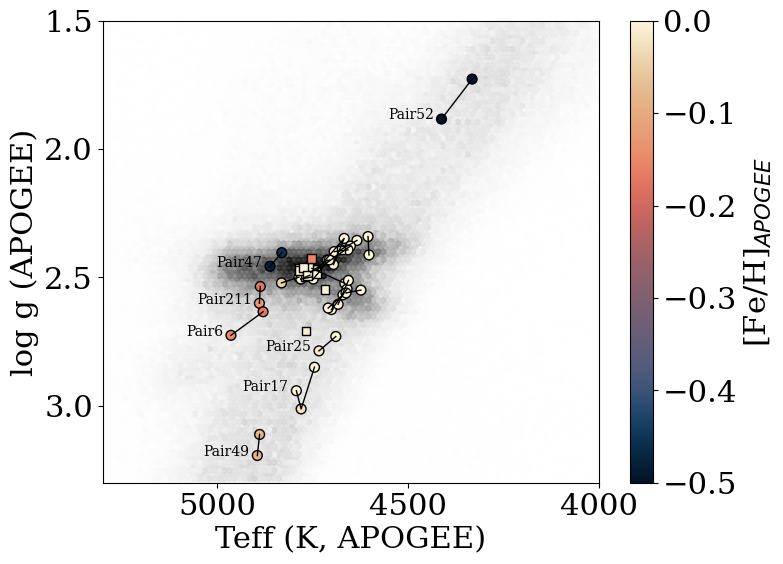}
    \caption{A Kiel diagram of our sample.  Doppelg{\"a}ngers are represented by circles connected with a line and colored by \textsl{APOGEE} DR17-reported [Fe/H].  Squares represent the observed reference M~67 stars.  The background presents field stars from the \textsl{APOGEE} DR17 survey colored by number density.  Pairs that lie at the edges of the distribution are labeled for interest.  Doppelg{\"a}ngers span the red giant branch with the majority occupying the red clump.  They display a range of metallicities, though the majority of doppelg{\"a}ngers have approximately Solar [Fe/H].}
    \label{fig:kiel}
\end{figure}

\subsection{Optical Spectra from the \textsl{Tull} Coud\'{e} Spectrograph on the McDonald Observatory 2.7m Telescope}
We obtained optical (3500 \AA~ \textless~$\lambda$ \textless~10,000 \AA), high-resolution (R$\sim$60,000) spectra of 25 \textsl{APOGEE} doppelg{\"a}ngers and 11 M~67 giant stars with the \textsl{Tull} Coud\'{e} Spectrograph on the McDonald Observatory 2.7m telescope using Slit 4 (\citealp{Tull1995}).  We aimed for signal-to-noise ratios (SNR) per pixel of at least 100 near the 5500 $\AA$ wavelength region to maximize abundance precision.  We summarize our observations in Table \ref{table:target_table}.  For simplicity, we refer to \textsl{APOGEE} doppelg{\"a}ngers by an arbitrary pair number, and each star in a pair is assigned a letter. "Pairs" 13 and 17 have four and three stars that qualify as doppelg{\"a}ngers, respectively.  Pairs 29 and 44 also share a star (the `a' component of both).

To illustrate that our final sample consists of apparently unrelated field stars, we present their on-sky positions in Figure \ref{fig:radec}.  Doppelg{\"a}ngers span a range of positions, on-sky separations, and parallaxes.  No pairs appear to be comoving as all have either large on-sky separations or significant differences in parallax, proper motion, and/or radial velocity.  In Figure \ref{fig:kiel}, we present a Kiel diagram for our sample of doppelg{\"a}ngers and M~67 reference stars.  The majority of our sample occupies the red clump, though a few pairs lie outside of the clump.  Most stars have Solar metallicity ([Fe/H] = 0 $\pm0.02$), though three pairs (Pairs 6, 49, and 211) have [Fe/H] $\sim$ -0.10, Pair 47 has [Fe/H] = -0.48, and Pair 52, our lowest surface gravity pair, has [Fe/H] = -0.58.

We note that star `a' in Pair 15 possesses a resolved stellar companion that we only discovered during data acquisition as it does not have enhanced RUWE and survived our cuts.  \citet{elbadry2021} reports that Pair 15a has a companion that is four magnitudes fainter and slightly bluer in color.  \textit{Gaia} reports a BP-RP color of 1.34 for the primary star and 0.98 for its companion.  The companion appears to be a lower mass main sequence star, so it is unlikely that it donated mass to the primary star and affected its photospheric abundances.  As such, we do not omit this pair from our study but do encourage caution when interpreting its abundances.

\begin{table}
\begin{tabular}{ccccc}
\hline
Gaia DR3 Source ID & Name & ObsDate & Exp & N$\rm _{exp}$ \\
 & & (YYYYMMDD) & (s) & \\
\hline
\hline
604923367432823808 & M~67 a & 20240119 & 1800 & 2 \\
604920202039656064 & M~67 b & 20240119 & 1800 & 2 \\
604919480485158784 & M~67 c & 20240119 & 1800 & 3 \\
.. & .. & .. & .. & .. \\
1509031094389703680 & Pair47 a & 20230530 & 1800 & 3 \\
1332869540410728448 & Pair47 b & 20230529 & 1800 & 2 \\
1393189710382298496 & Pair49 a & 20230531 & 1800 & 2 \\
2106262308540128000 & Pair49 b & 20230531 & 1800 & 2 \\
1313275448633710080 & Pair50 a & 20230529 & 1200 & 1 \\
1414698013247715840 & Pair50 b & 20230529 & 1800 & 3 \\
1489165668053970176 & Pair52 a & 20230613 & 1800 & 2 \\
1201468623762697856 & Pair52 b & 20230613 & 1800 & 4 \\
2593796720752077568 & Pair6 a & 20240119 & 1800 & 1 \\
173154016016909568 & Pair6 b & 20240119 & 1800 & 6 \\
\hline
\end{tabular}
\caption{A summary of our observations.}\label{table:target_table}
\end{table}

After acquisition, the raw \textsl{Tull} Coud\'{e} Spectrograph data are initially processed using the \textsl{Tull} Coud\'{e} Spectrograph Data Reduction Pipeline (\texttt{TSDRP}\footnote{\url{https://github.com/grzeimann/TSDRP}}). This pipeline performs essential calibration and extraction steps, including bias subtraction, trace identification, scattered light subtraction, wavelength calibration, flat-field correction, cosmic ray rejection, and spectral extraction for each spectral order. Additionally, \code{TSDRP} provides deblazing, continuum normalization, and order combination to produce a single, fully processed spectrum.  Radial velocities are determined for each spectrum using \code{iSpec}, and spectra are shifted to rest air wavelengths.  Finally, sub-exposures of the same object are coadded by spline interpolating fluxes onto a shared wavelength grid and summing them together, weighing each flux array by the SNR of the subexposure.  This method is more effective than median coadding when observing conditions (e.g., clouds, seeing) vary throughout the exposures \citep[see discussion and treatment of this problem in][]{Hogg2024}.

\section{Method}\label{sec:methods}

In this work, we require high abundance precision to enable the detection of potentially subtle chemical differences between pairs of stars.  Precise stellar parameters are critical for achieving high abundance precision.  Due to the high SNR of the APOGEE spectra (SNR \textgreater 300 for the doppelgangers), \textsl{APOGEE} reports precise stellar parameters for our sample (mean $\rm T_{eff}$ and log g uncertainties of 8 K and 0.02 dex, respectively).  Thus, we adopt APOGEE's $\rm T_{eff}$ and log g values and uncertainties for this work.  Our analysis is entirely differential in nature, so precise \textit{differences} in stellar parameters among stars are the priority.  As such, concerns about global offsets between absolute \textsl{APOGEE}-reported stellar parameters and absolute stellar parameters derived from the optical \citep[e.g.,][]{Nandakumar2022, Hegedus2023} are not relevant in this differential context.

Though we adopt APOGEE $\rm T_{eff}$ and log g, we must still determine microturbulence (v$_{\rm micro}$) and spectral broadening from our optical spectra.  Though v$_{\rm micro}$ is reported by APOGEE, we find that adopting it for our spectra can lead to trends between the abundances of Fe I lines and their reduced equivalent widths.  This can inflate the line-to-line abundance scatter and thus affect our abundance precision.  As a result, we elect to measure our own v$_{\rm micro}$.  We use Brussels Automatic Code for Characterizing High accUracy Spectra \citep[BACCHUS,][]{BACCHUS} to determine v$_{\rm micro}$, spectral broadening, and individual elemental abundances for our optical spectra.  Additionally, we investigate whether \textsl{APOGEE}-reported $\rm T_{eff}$ and log g are consistent with results returned by BACCHUS using the Fe ionization-excitation method (see description in next section).  The results of this test are presented in the Appendix.

\subsection{BACCHUS} \label{subsec:BACCHUS}
We describe BACCHUS briefly in the following paragraphs and point readers to  \citet{Hawkins2016}, \citet{Nelson2021}, \citet{Hayes2022}, and the official manual\footnote{BACCHUS Manual: \url{https://drive.google.com/file/d/1VShSwA5M21q2pSSxLxc9AnoA19ixu-eV/view}} for a more detailed description.  BACCHUS is a spectral synthesis and fitting tool designed for high resolution data.  It synthesizes spectra using the radiative transfer code TURBOSPECTRUM \citep{Plez2012} adopting the MARCS model atmosphere grid \citep{Gustafsson2008} and assuming one-dimensional local thermodynamic equilibrium (1D LTE).  We adopt version 5 of the \textit{Gaia}-ESO linelist for our atomic transition data \citep{Heiter2021} and combine molecular transition data from numerous sources: CH from \citealt{Masseron2014}, C2, CN, OH, and MgH from T. Masseron, private communication, SiH from \citealt{Kurucz1992}, and TiO, FeH, and ZrO from B. Pelz, private communication.  

Stellar parameters for which BACCHUS can fit are $T_{\rm eff}$, log g, [M/H], microturbulence (v$_{\rm micro}$), and spectral broadening, called the "convolution" parameter, which includes broadening due to instrumental resolution, macroturbulence, and projected stellar rotation.  BACCHUS uses traditional Fe ionization-excitation balance to solve for stellar parameters in an iterative fashion.  The process begins with the user's initial guess of the stellar parameters which are used to interpolate over the MARCS grid and produce a starting model atmosphere.  Initial abundances of individual Fe lines are determining by minimizing the $\chi^2$ between the observed spectral line segment and an interpolation across five synthetic absorption lines with varying Fe abundance.  Stellar parameters are then solved for simultaneously using an iterative process that adjusts the input model atmosphere in light of the last iteration's results.  When solving for v$_{\rm micro}$, the only parameter that we re-determine from the optical spectra, BACCHUS converges when there is a null trend between the abundances and reduced equivalent widths of Fe I lines. With BACCHUS, we obtain average v$_{\rm micro}$ uncertainties of 0.05 km/s $\pm$ 0.02 km/s. To solve for the convolution parameter, BACCHUS iterates through possible values until it achieves an agreement between Fe line abundances determined using the line intensity and equivalent width methods (see next paragraph).

With $T_{\rm eff}$, log g, and [M/H] from \textsl{APOGEE} and v$_{\rm micro}$ and spectral broadening from BACCHUS, we determine abundances in 21 elements that span five nucleosynthetic families: the $\alpha$ (Mg, Ca, Si), odd-Z (Na, Al), Fe-peak (Fe, Sc, Ti, V, Cr, Co, Ni, Cu\footnote{Cu is also considered a neutron-capture element \citep[e.g.,][]{Baratella2021}.}, Zn), slow neutron-capture process (Y, Zr, Ba, La, Ce), mixed (Nd), and rapid neutron-capture process (Eu) elements.  We omit the light elements as we lack access to reliable C, N, and O lines. BACCHUS uses four methods to determine a line's abundance: $\chi^2$ minimization, synthetic equivalent widths, spectral synthesis, and line intensity.  Each method starts in the same way.  After interpolating across the MARCS model grid to obtain an atmosphere parameterized by the $T_{\rm eff}$, log g, [M/H], microturbulence, and (v$_{\rm micro}$) determined or set in the previous step, five lines are synthesized: one line with the expected abundance assuming a Solar abundance pattern scaled to the input model metallicity and four additional lines with -0.6, -0.3, +0.3, and +0.6 dex relative to the expected abundance.  The $\chi^2$ method computes the $\chi^2$ between the observed and five synthetic spectra and adopts the abundance that minimizes the polynomial fit to the $\chi^2$ vs. abundance trend.  The equivalent width method measures the equivalent width of the five synthetic lines and uses interpolation to find the abundance that minimizes the difference between the synthetic and observed equivalent width.  The spectral synthesis method is similar to $\chi^2$ but instead finds the abundance solution that minimizes the difference between the synthetic and observed spectrum.  Finally, the line intensity method is similar to the equivalent width method but considers line depth (the average flux of the five points nearest the line center) instead of equivalent width. The equivalent width and line intensity methods will only agree when a suitable spectral broadening parameter is assumed.  When determining abundances, we only consider abundances from lines where all four methods return unflagged results (see description of flags in \citealp{Hayes2022}) and the $\chi^2$ and equivalent width methods return consistent abundances within 0.1 dex.

\subsubsection{Line-by-line Differential Abundance Analysis}
To achieve high precision abundances, we perform a line-by-line differential analysis to determine abundance differences among stars in doppelg{\"a}nger or open cluster pairs.  Line-by-line differential analysis is an effective way of minimizing abundance uncertainties and has been used widely for this purpose (again, e.g., \citealp{Melendez2012, Yong2013, Melendez2014, Bedell2018, Hawkins2020, Liu2021, Nelson2021, Yong2023, Liu2024}, and references in \citealp{Nissen2018}).  In standard (non-differential) abundance analysis, the abundance difference between two stars is determined in two steps: 1) adopting the final abundance of a star as the mean abundance derived from all well-measured lines, and 2) subtracting the final abundance of one star from another.  This method is subject to sources of uncertainty that the line-by-line differential method bypasses.  Due to poorly constrained log \textsl{gf}s\footnote{The log \textsl{gf} of a line is the product of the statistical weight of the lower level of the transition (\textsl{g}) and the oscillator strength (\textsl{f}), see discussion in \citealp{Gray2008}.} and systematics within the data reduction process (e.g., consistently underestimating the continuum near broad line regions, poorly addressing imperfections in certain regions of the CCD), lines of the same element across the same stellar spectrum will return inconsistent abundances.  This inconsistency enlarges the final abundance uncertainty. In line-by-line differential abundance analysis, however, this line-to-line inconsistency becomes unimportant:  abundance differences are measured independently for each individual line shared by both spectra, and the mean line-by-line abundance difference is adopted as the final abundance difference. In this way, uncertain log \textsl{gf}s and systematic imperfections in the data reduction process will affect \textit{both} lines, and their impact will cancel out.  See \citet{Gray2008} for more on this effect.


\begin{table}
\begin{tabular}{ccc}
\hline
Element & Ion & Central Wavelength \\
 &  & (\AA)\\
 \hline
 \hline
Na & 1 & 4751.8 \\
Ti & 1 & 4758.1 \\
Ti & 1 & 4759.3 \\
Ce & 2 & 4773.9 \\
.. & .. & .. \\
Fe & 1 & 5364.9 \\
Fe & 1 & 5365.4 \\
Ti & 1 & 5366.6 \\
Fe & 1 & 5379.6 \\
Fe & 1 & 5383.4 \\
.. & .. & .. \\
Fe & 1 & 8621.6 \\
Al & 1 & 8773.9 \\
Ti & 1 & 8778.7 \\
Fe & 1 & 8846.7 \\
\hline
\end{tabular}
\caption{A list of the absorption lines used for abundance determination. Note that BACCHUS determines abundances with line-by-line spectral synthesis.  As such, the above central wavelengths describe the center of the spectral window used in line-by-line abundance determination.  Atomic data for the synthesis of each window are adopted from version 5 of the \textit{Gaia}-ESO linelist.  This list encompasses all possible spectral regions considered for abundance determination, though the specific subset of regions used for each star varies depending on individual spectral quality.}\label{tab:lineselect}
\end{table}

\begin{table*}
\begin{tabular}{ccc|cccc}
\hline
& & & \multicolumn{4}{c}{$\Delta \log \varepsilon(X)_\lambda$ given} \\
Element & Ion & Line (\AA) & $\Delta \rm T_{eff}$ = 100 K & $\Delta \rm log g$ = 0.10 & $\Delta \rm [M/H]$ = 0.10 & $\Delta \rm v_{micro}$ = 0.10 km/s \\
\hline
\hline
Na & 1 & 4751.8 & 0.03 & 0.04 & 0.04 & 0.03 \\
Ti & 1 & 4758.1 & 0.07 & 0.03 & 0.08 & 0.12 \\
Ti & 1 & 4759.3 & 0.06 & 0.04 & 0.03 & 0.07 \\
Ce & 2 & 4773.9 & 0.03 & 0.06 & 0.06 & 0.03 \\
.. & .. & .. & .. & .. & .. & ..\\
Fe & 1 & 5364.9 & 0.1 & 0.04 & 0.04 & 0.11 \\
Fe & 1 & 5365.4 & 0.08 & 0.04 & 0.08 & 0.18 \\
Ti & 1 & 5366.6 & 0.11 & 0.05 & 0.07 & 0.08 \\
Fe & 1 & 5379.6 & 0.08 & 0.03 & 0.04 & 0.11 \\
Fe & 1 & 5383.4 & 0.13 & 0.05 & 0.03 & 0.07 \\
.. & .. & .. & .. & .. & .. & ..\\
Fe & 1 & 8621.6 & 0.09 & 0.06 & 0.01 & 0.15 \\
Al & 1 & 8773.9 & 0.11 & 0.02 & 0.03 & 0.05 \\
Ti & 1 & 8778.7 & 0.07 & 0.04 & 0.1 & 0.04 \\
Fe & 1 & 8846.7 & 0.02 & 0.03 & 0.03 & 0.05 \\
\hline
\end{tabular}
\caption{The sensitivity of each line's derived elemental abundance ($\log \varepsilon(X)_\lambda$) in response to perturbing the adopted stellar atmosphere by $\Delta \rm T_{eff}$ = 100 K, $\Delta \rm log g$ = 0.10, $\Delta \rm [M/H]$ = 0.10, and $\Delta \rm v_{micro}$ = 0.1 km/s.  This table is used to assess the impact of stellar parameter uncertainties on final differential abundance uncertainties.  While each element responds differently to changes in stellar parameters, the light, $\alpha$, odd-Z, and Fe-peak elements are typically more sensitive to changes in $\rm T_{eff}$ while the neutron-capture elements are more sensitive to changes in log g.}\label{tab:abund_sens}
\end{table*}

In this work, each doppelg{\"a}nger's abundance difference is determined by taking the average line-to-line abundance difference measured from all lines shared by the two stellar spectra, weighed by each line's differential abundance uncertainty. To determine each line's differential abundance uncertainty, we follow the formalism described in Appendix A of \citet{Ji2020}.  In short, we estimate the sensitivity of each line's measured abundance difference to changes in $\Delta \rm T_{eff}$, $\Delta$ logg, $\Delta$ [M/H], and $\rm \Delta v_{micro}$ among the two spectra using five representative doppelg{\"a}nger pairs in our sample (Pairs 1, 13, 29, 49, and 52).  This is accomplished by measuring the average line-by-line abundance differences among pairs where one star's $\rm T_{eff}$, logg, [M/H], or $\rm v_{micro}$ was perturbed by 100K, 0.1, 0.1, and 0.1 km/s, respectively.  We present the average differential abundance sensitivities in Table \ref{tab:abund_sens}.  Note how different elements and lines show different sensitivities to changing stellar parameters.  From here, we can derive empirical relationships between changes in  $\Delta T_{\rm eff}$, $\Delta$ log g, $\Delta$ [M/H], or $\rm \Delta v_{micro}$ and $\rm \Delta \log \varepsilon(X)_{\lambda}$, the abundance difference in line $\lambda$ of element X.  We apply these relationships to the stellar parameter uncertainties (again, using the method of \citealt{Ji2020}) to determine line-by-line differential abundance uncertainties.  With differential abundance uncertainties estimated for each line, we can now estimate differential abundances and associated uncertainties for each element.  Again, we do this by taking the average abundance difference across all lines of the element \textit{X} weighted by the inverse variance of the individual differential abundance uncertainty of each line $\lambda$. As in \citet{Ji2020}, the final adopted differential abundance uncertainty takes into account stellar parameter-driven differential abundance uncertainties and the line-to-line differential abundance scatter.

\section{Results}\label{sec:results}

\subsection{\textsl{APOGEE} DR17 Doppelg{\"a}nger Rate}
\begin{figure}
    \centering
    \includegraphics[width=.5\textwidth]{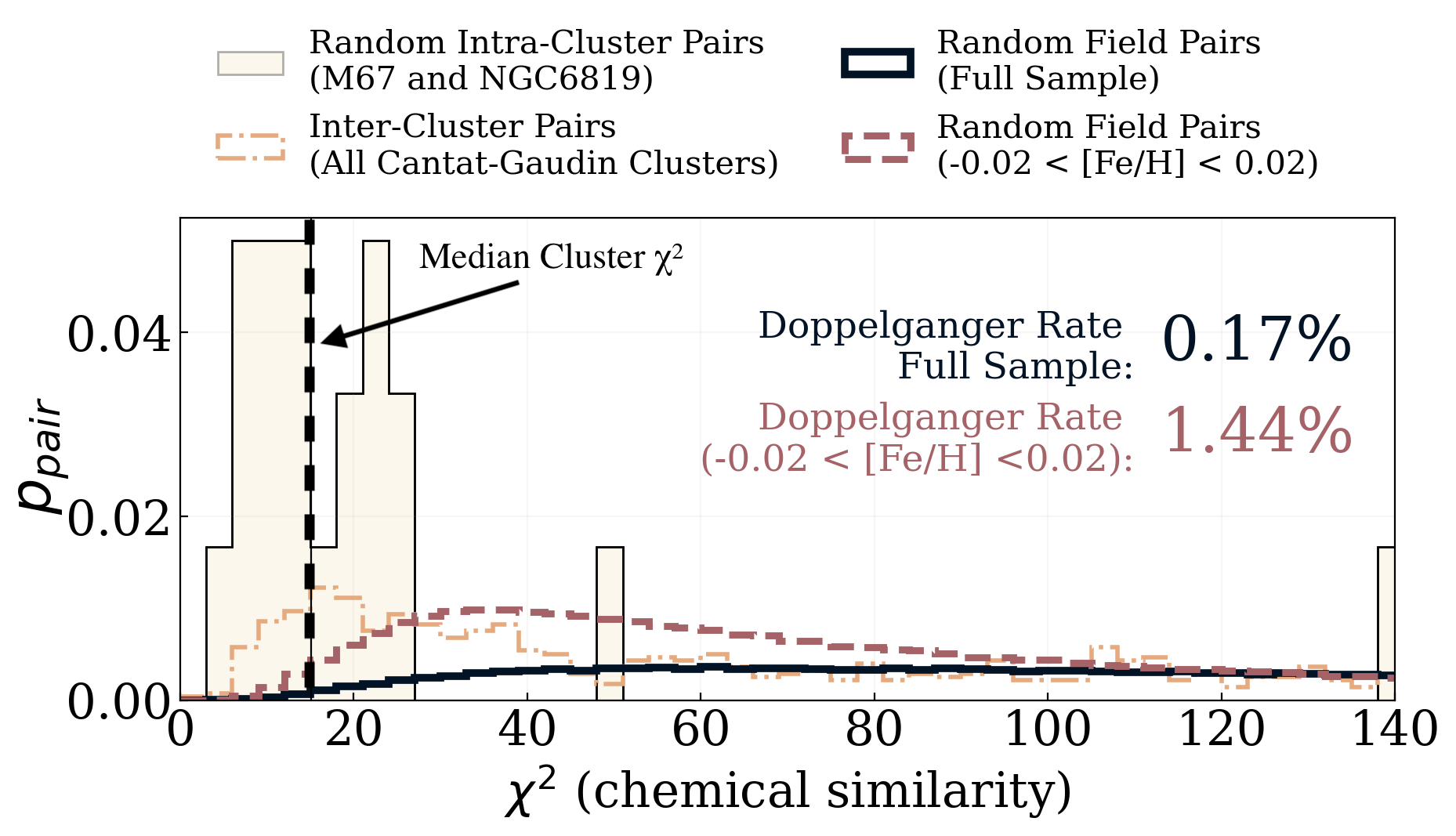}
    \caption{Our measurement of the \textsl{APOGEE} DR17 doppelg{\"a}nger rate, defined as the rate at which randomly drawn pairs of field stars are as chemically similar as stars born together.  This measurement is conducted by identifying the rate at which field stars possess $\chi^2$ values (a metric of chemical similarity, see Equation \ref{eq:chisq}) less than or equal to the median $\chi^2$ value of stellar pairs drawn from within open clusters. The solid black and dashed red curves present the distribution of $\chi^2$ values for randomly drawn pairs of \textsl{APOGEE} field stars spanning the full [Fe/H] range and -0.02 < [Fe/H] < 0.02, respectively.  The $\chi^2$ distribution for intracluster pairs is shown in yellow.  As in N18, we also present the $\chi^2$ distribution for inter-cluster (across different clusters) pairs with the orange dot-dash outline.  Doppelg{\"a}ngers exist to the left of the median intracluster $\chi^2$ value (vertical line).  We measure a doppelg{\"a}nger rate of 0.17\% for the full DR17 sample and 1.44\% for the narrow [Fe/H] range, comparable to similar measurements in DR13 and 16.}
    \label{fig:apogee_dr}
\end{figure}

This work studies the chemical similarity of \textsl{APOGEE}-identified ``chemical doppelg{\"a}ngers" from the optical, high resolution point of view.  To do this, we first identify doppelg{\"a}ngers for optical follow-up.  In doing so, we conduct the first measurement of the \textsl{APOGEE} DR17 doppelg{\"a}nger rate (Figure \ref{fig:apogee_dr}).  The doppelg{\"a}nger rate is a measure of the chemical diversity of stars in a population and places important constraints on Galactic chemical evolution. We find an \textsl{APOGEE} DR17 doppelg{\"a}nger rate of 0.17\% when considering the full high-fidelity \textsl{APOGEE} sample (see Section \ref{sec:apogee_selection}) and 1.44\% for stars -0.02 \textless [Fe/H] \textless 0.02.  Both of these values agree well with previous measurements in DR 13 \citep{Ness2018} and DR 16 \citep{deMijolla2021}.  The median intracluster pairwise $\chi^2$ value (see Equation \ref{eq:chisq}) is $\sim$15, just lower than the number of elements considered in computing the metric (17).  This indicates that the reference open clusters (M~67 and NGC 6819) are generally chemically homogeneous at or below the abundance uncertainty level in the elements considered according to \textsl{APOGEE} DR17.  

\subsection{Line-by-line Differential Abundances of Stellar Pairs}

\subsubsection{Chemical Homogeneity of Reference Open Cluster M~67} \label{sec:m67}
According to our adopted definition, chemical doppelg{\"a}ngers are random pairs of disk stars that are as chemically similar as stars born together.  To confirm the ``doppelg{\"a}nger" status of our sample, we must first constrain the chemical similarity of stars born together.  We pair our observed M~67 stars into all unique combinations that satisfy the $\rm T_{eff}$ and logg similarity requirements of our initial doppelg{\"a}nger selection ($\Delta \rm T_{eff} < 50 K, \Delta log g < 0.1 dex$).  This results in 24 unique open cluster pairs.  We determine the average uncertainty-corrected abundance differences of these pairs (computed as the quadratic difference between the abundance difference and the associated uncertainty)  and report them in Table \ref{tab:oc}.  We find that M~67 pairs on average differ by less than 0.02 dex (with the exception of Na, which shows an average abundance difference of 0.033) in [Fe/H] and [X/Fe], consistent with previous investigations of the cluster's chemical homogeneity \citep[e.g.,][]{Bovy2016b, Ness2018, Sinha2024, GALAHDR4OCS}.


\hspace{-1cm}
\begin{table*}
\centering
\begin{tabular}{c|ccc|ccc}
\hline
& \multicolumn{3}{c|}{M67} & \multicolumn{3}{c}{Doppelgangers} \\
 & Average $\Delta$[X/Fe]\footnote{Except for Fe which is reported as [Fe/H]}  & Average $\Delta$[X/Fe]$\rm ^{a}$  & Average $\Delta$[X/Fe]$\rm ^{a}$ & Average $\Delta$[X/Fe]$\rm ^{a}$  & Average $\Delta$[X/Fe]$\rm ^{a}$  & Average $\Delta$[X/Fe]$\rm ^{a}$\\
El & (corrected) & (uncorrected) &  Uncertainty & (corrected) & (uncorrected) & Uncertainty \\
\hline
\hline
Fe & 0.018 & 0.018 & 0.006 & 0.017 & 0.017 & 0.006 \\
Na & 0.033 & 0.044 & 0.022 & 0.012 & 0.026 & 0.025 \\
Mg & 0.013 & 0.029 & 0.024 & 0.017 & 0.031 & 0.027 \\
Al & 0.005 & 0.014 & 0.012 & 0.013 & 0.022 & 0.012 \\
Si & 0.015 & 0.022 & 0.009 & 0.013 & 0.023 & 0.01 \\
Ca & 0.002 & 0.015 & 0.02 & 0.001 & 0.013 & 0.022 \\
Sc & 0.009 & 0.019 & 0.015 & 0.02 & 0.031 & 0.016 \\
Ti & 0.016 & 0.024 & 0.009 & 0.006 & 0.015 & 0.01 \\
V & 0.021 & 0.029 & 0.011 & 0.009 & 0.018 & 0.012 \\
Cr & 0.019 & 0.027 & 0.014 & 0.016 & 0.027 & 0.015 \\
Co & 0.005 & 0.015 & 0.013 & 0.012 & 0.021 & 0.015 \\
Ni & 0.007 & 0.018 & 0.016 & 0.009 & 0.019 & 0.017 \\
Cu & 0.0 & 0.018 & 0.039 & 0.007 & 0.022 & 0.043 \\
Zn & 0.016 & 0.052 & 0.082 & 0.004 & 0.040 & 0.090 \\
Y & 0.001 & 0.009 & 0.014 & 0.022 & 0.032 & 0.015 \\
Zr & 0.003 & 0.021 & 0.02 & 0.033 & 0.044 & 0.023 \\
Ba & 0.0 & 0.012 & 0.019 & 0.048 & 0.062 & 0.022 \\
La & 0.013 & 0.035 & 0.038 & 0.04 & 0.057 & 0.029 \\
Ce & 0.012 & 0.03 & 0.037 & 0.031 & 0.044 & 0.028 \\
Nd & 0.009 & 0.022 & 0.019 & 0.03 & 0.042 & 0.021 \\
Eu & 0.006 & 0.016 & 0.019 & 0.016 & 0.029 & 0.02 \\
\hline
\end{tabular}
\caption{Mean abundance differences (both uncertainty corrected and uncorrected, see Section \ref{sec:m67}) and mean abundance difference uncertainties for the M~67 and doppelg{\"a}nger pairs.}\label{tab:oc}
\end{table*}

\subsubsection{Differential Abundances of \textsl{APOGEE}-identified Chemical Doppelg{\"a}ngers}
\begin{figure*}
    \centering
    \includegraphics[width=\textwidth]{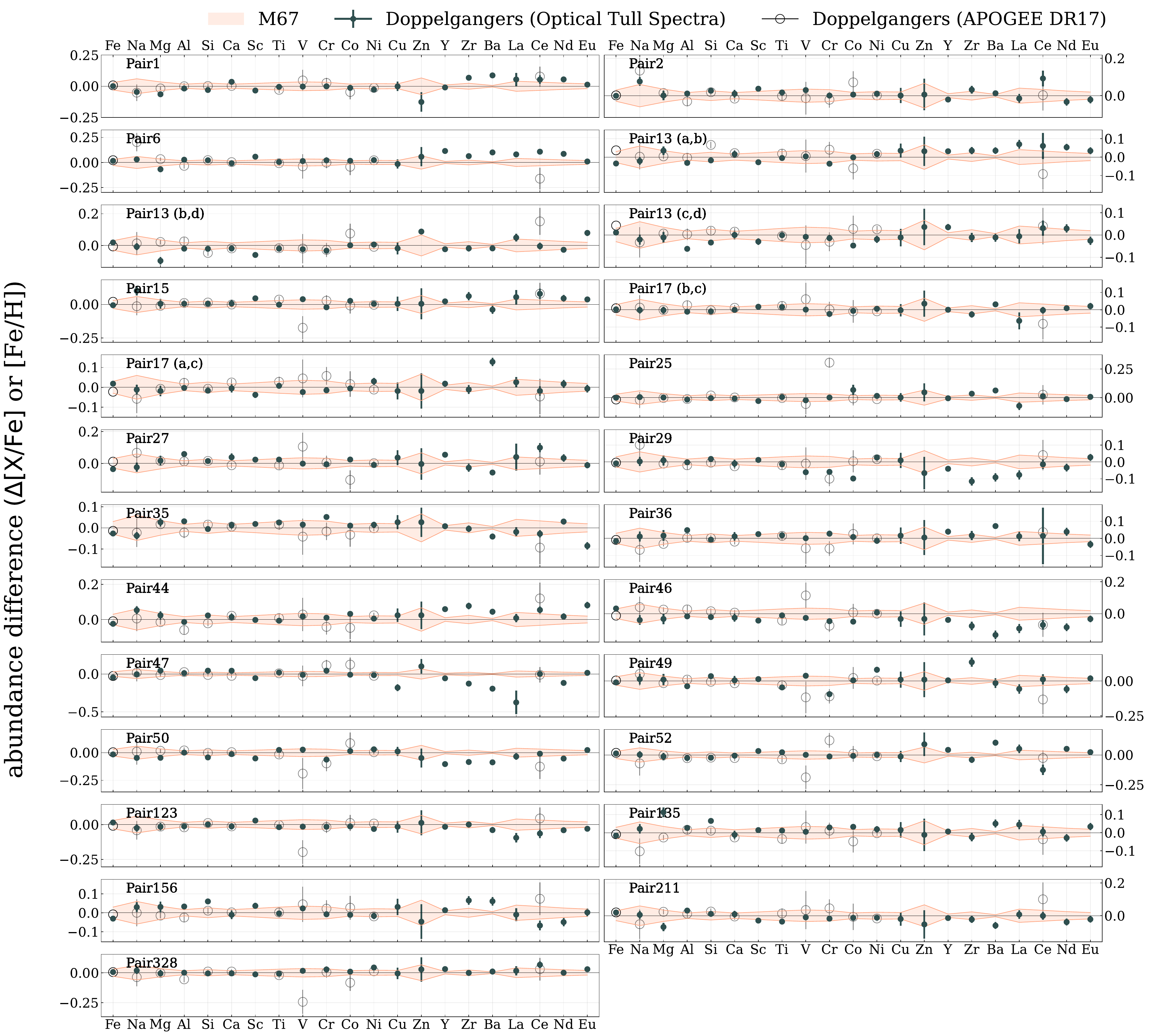}
    \caption{Differences in [X/Fe] (or, for Fe, [Fe/H]) for each \textsl{APOGEE}-identified chemical doppelg{\"a}nger pair (filled circles) determined from our optical \textsl{Tull} spectra.  Open circles indicate the equivalent \textsl{APOGEE} DR17-reported values where available.  Orange fill indicates the standard deviation in abundance difference among M67 pairs.  Doppelgangers are generally highly similar in the elements used in their initial selection (those with open circles that \textsl{APOGEE} can measure) but can differ significantly (beyond what is typical among M~67 pairs) in the neutron-capture elements that \textsl{APOGEE} cannot access.}
    \label{fig:bigdiff}
\end{figure*}

With constraints on the abundance similarity of our reference M~67 pairs, we can now assess the abundance similarities of our doppelg{\"a}nger pairs.  Given their selection, pairs should show small abundance differences comparable to those of the M~67 pairs in the \textsl{APOGEE}-reported elements.  We present the results of our line-by-line differential abundance analysis in Figure \ref{fig:bigdiff}.  According to our definition, stars are considered to be doppelg{\"a}nger in an element when their abundance difference (filled circles) is less than the typical abundance differences among open cluster pairs.  For this portion of our analysis, we use the standard deviation in the abundance differences among M 67 pairs (orange fill) as a reference point. Our optical analysis confirms that \textsl{APOGEE}-identified chemical doppelg{\"a}ngers are indeed generally doppelg{\"a}ngers in the \textsl{APOGEE}-measured elements on which they were selected (Na through Fe in the periodic table).  However, there are a few exceptions to this.  

Pairs 2 and 15 show 0.08-0.10 dex [Na/Fe] differences that exceed the typical differences in M67 pairs.  \textsl{APOGEE} also detects a significant [Na/Fe] difference in Pair 2 but with larger uncertainties.  [Al/Fe] can also differ slightly (\textless 0.03 to 0.06 dex) for several pairs (Pairs 13's c and d components, 27, 36, 49, and 156).  These differences are small but significant because they are larger than the average for M~67 pairs.  [Mg/Fe] differs by 0.05 to 0.06 dex in Pairs 6, 13 (b and d), and 135, exceeding M~67's 0.03 dex standard deviation in abundance difference.  Si and Ca also differ subtly between pairs, with Pair 135 showing the largest difference (0.05 dex for Si).  Several pairs (6, 13 b and d components, 15, 46, 47, and 50) show 0.04 to 0.06 dex differences in Sc but no other Fe-peak elements.  Pair 328 shows a $\sim$0.07 dex difference in [Ni/Fe] but again no significant differences in the remaining Fe-peak elements.  Pair 49 shows significant Ni and Cr differences but no such differences in other Fe-peak elements.  Finally, Co differs slightly among Pairs 13 (c and d), 15, 46, and 135.  Barring these exceptions, our optical analysis generally confirms the "doppelg{\"a}nger" status of these stars in the lighter (Z \textless 29) elements on which they were selected, and our differential results agree well with \textsl{APOGEE}'s.

Next, we will discuss the differential abundance results for the elements newly measured in the optical.  Zn and Cu are elements either missing in or not well-measured by \textsl{APOGEE} and thus not used in the initial selection of doppelg{\"a}ngers.  However, we can measure two Zn I lines ($\lambda$ 4722.159, 4810.54 \AA) up to four Cu I lines in our optical spectra ($\lambda$ 5105.5, 5218.2,5220.1, and 5700.2 \AA).  Pairs 1, 13 (b and d), 47, and 52 show 0.09 to 0.13 dex differences in [Zn/Fe], though we highlight the significant 0.07 dex [Zn/Fe] spread in M67 for comparison and the 0.08 dex average abundance uncertainty). Pairs 27 and 47 are the only pairs to show significant (0.04 and 0.18 dex, respectively) differences in [Cu/Fe] compared to the 0.02 dex standard deviation among M 67 pairs' abundance differences.


Finally, we can assess the neutron-capture (Zr, Y, La, Ce, Nd, and Eu) elemental abundance similarities of our doppelg{\"a}nger sample.  None of the 25 pairs can be considered doppelg{\"a}nger in all neutron-capture elements.  All doppelg{\"a}ngers show abundance differences in at least one neutron-capture element.  Fifteen pairs show differences in one or both light \textit{s}-process elements [Y/Fe] and [Zr/Fe] that exceed the typical abundance spread within M~67.  Nineteen, eleven, and twelve pairs differ significantly in heavy \textit{s}-elements Ba, La, and Ce, respectively, beyond the typical difference among M67 pairs (up to 0.380 $\pm$ 0.150 dex for [La/Fe] in Pair 47).  Mixed element Nd (see Section \ref{sec:discussion} for more on mixed elements) differs in ten pairs. Finally, Eu differs significantly among five pairs (Pairs 13 b\&d, 15, 35, 36, and 44).

\begin{figure*}
    \centering
    \includegraphics[width=.9\textwidth]{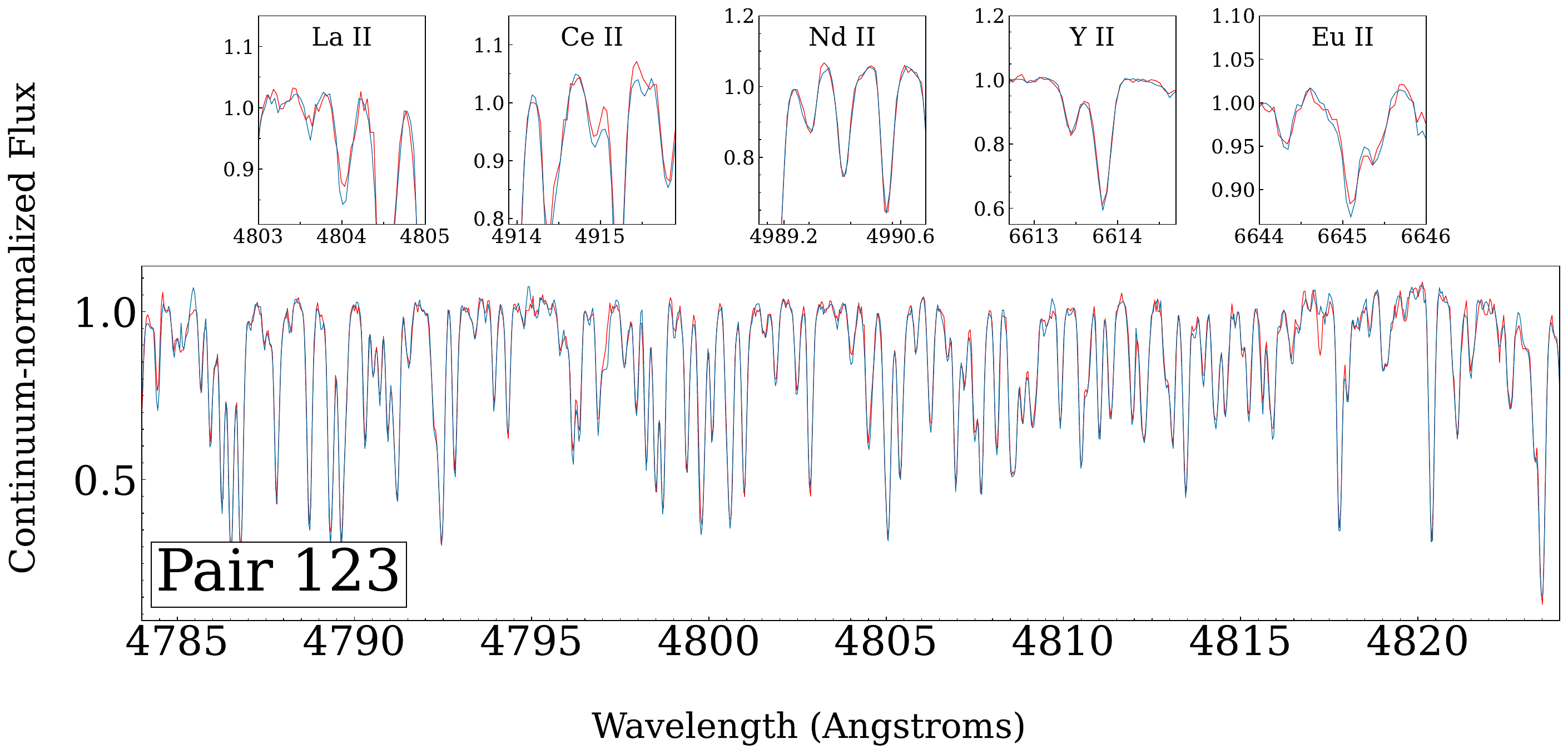}
    \caption{\textsl{Tull} spectra of the a (red) and b (blue) components of Pair 123.  This pair differs detectably in [La/Fe] and [Ce/Fe] but not [Nd, Y, or Eu/Fe], and signs of this can be seen in the spectral zoom-ins at the top.  Beside these differences, doppelg{\"a}ngers generally share remarkably similar optical spectra.}
    \label{fig:spec}
\end{figure*}

There are instances where doppelg{\"a}ngers differ in a subset of the elements within a nucleosynthetic family but not the rest.  For example, in Pair 123, the stars differ significantly in La and Ce but share indistinguishable compositions in the remaining neutron-capture elements.  We visually inspect all doppelg{\"a}nger spectra to confirm that abundance differences reflect line flux differences.  We show an example comparison for Pair 123, whose spectra display generally indistinguishable Y, Nd, and Eu lines but flux differences around the La and Ce lines (Figure \ref{fig:spec}).

\begin{figure*}
    \centering
    \includegraphics[width=0.9\linewidth]{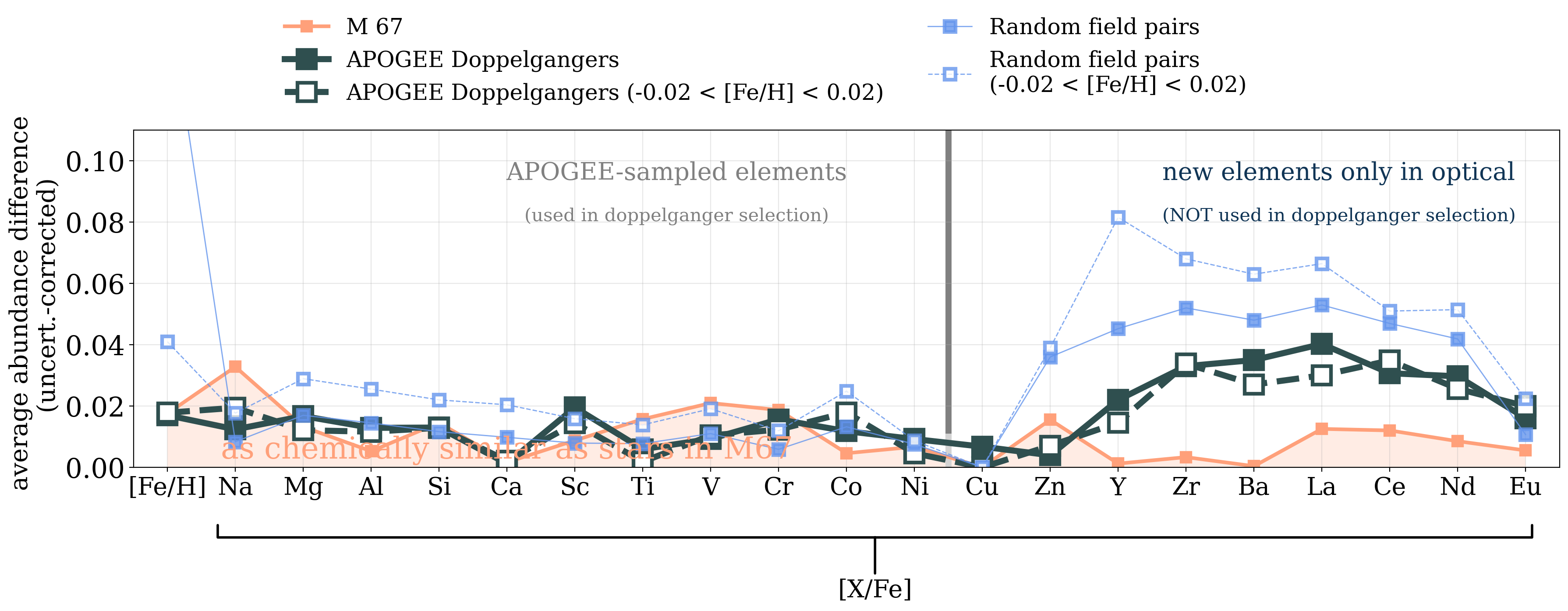}
    \caption{Average uncertainty-corrected abundance differences among all doppelg{\"a}ngers (filled dark green squares) and just those with Solar metallicity (open green squares).  For comparison, we include the equivalent for M~67 pairs (orange squares) and random, chemically unrelated field pairs (blue filled and open squares representing all and exclusively Solar metallicity field pairs).  The position of the dark green squares with respect to the orange ones indicates the distinguishing power of each element among \textsl{APOGEE}-identified chemical doppelg{\"a}ngers.  This figure illustrates that Y, Zr, Ba, La, Ce, and Nd can typically distinguish between \textsl{APOGEE}-identified doppelg{\"a}ngers.}
    \label{fig:summary}
\end{figure*}

We summarize our differential analysis results in Figure \ref{fig:summary}, where we present the average uncertainty-corrected abundance differences among our doppelg{\"a}nger pairs.  Again, we define the uncertainty-corrected abundance difference to be the quadratic difference between the measured abundance and the associated uncertainty.  This corrected form takes into account uncertainties and allows us to determine average abundance differences without adding scatter due to large uncertainties.  To contextualize these differences, we compare them to the average uncertainty-corrected abundance differences among M~67 pairs, a reference measure for the chemical similarity of stars born together.  We also compare them to the opposite extreme, the chemically similarity of stars that were \textit{not} born together, by determining the average uncertainty-corrected abundance differences among \textit{random field stars} that are chemically unrelated. This comparison sample of random field pairs is created by recycling our doppelg{\"a}nger sample but pairing stars randomly and excluding combinations of doppelg{\"a}ngers.  We are unable to impose our strict $\Delta \rm T_{eff}~and~logg$ requirements for the random field pairs due to a lack of non-doppelg{\"a}nger combinations that satisfy them, but we require that random pairs have $\Delta \rm T_{eff} < 100 K~and~\Delta logg < 0.15~dex$.  The standard deviation in $\Delta$ [X/Fe] for our doppelg{\"a}nger sample captures the typical doppelg{\"a}nger star-to-star difference in each element.  When compared to the standard deviation of M~67 pairs, this quantity also informs the distinguishing power of each element among \textsl{APOGEE}-identified chemically similar stars. For example, when doppelg{\"a}ngers' differential abundance spread in an element is less than or equal to that of M~67 pairs, then the element does not typically distinguish \textsl{APOGEE}-identified chemical doppelg{\"a}ngers.  When the typical abundance difference exceeds that of M~67, this element can be used to distinguish between \textsl{APOGEE}-identified doppelg{\"a}ngers.

Figure \ref{fig:summary} further confirms that in general, \textsl{APOGEE}-identified chemical doppelg{\"a}ngers indeed appear to be doppelg{\"a}ngers in the elements used to select them, though [Al, Si, and Sc/Fe] can differ slightly (\textless 0.005 dex on average) among pairs beyond the typical differences found in M~67 pairs.  However, doppelg{\"a}ngers can show large abundance differences in the elements that \textsl{APOGEE} does not measure (or measures imprecisely).  Our results suggest that \textsl{APOGEE}-identified doppelg{\"a}ngers can typically be distinguished by their differences in one or more \textit{s}-process elements: the light \textit{s}-process elements Y and Zr, the heavy \textit{s}-process elements Ba, La, and Ce, and/or mixed element Nd.  Within our sample of doppelg{\"a}ngers, Y, Zr, Ba, La, Ce, and Nd tend to differ by an amount 0.02-0.05 dex greater than that among M~67 pairs.  Interestingly, \textit{r}-process element Eu shows comparatively smaller differences among doppelgangers, exceeding those of M~67 pairs by 0.01 dex.  [Eu/Fe] is thus less effective at distinguishing doppelg{\"a}ngers at the current precision of 0.02 dex.  Furthermore, weak \textit{s}-process elements Cu and Zn do not typically distinguish among doppelg{\"a}ngers.

\subsection{Correlations Between Chemical Similarity and Other Stellar Attributes}

Our results so far have indicated that stars can appear to be chemically indistinguishable in \textsl{APOGEE} but differ significantly in some elements according to their high resolution optical spectra.  It is of interest to understand why some doppelg{\"a}ngers display abundance differences while others do not.  Searching for relationships between doppelg{\"a}ngers' abundance similarity and their similarity in other parameters may provide insights.  Covariances between abundance similarity in elements produced through the same nucleosynthetic channel can validate our abundance determination method and provide insights into the nucleosynthetic origins of these abundance differences.  For example, if doppelg{\"a}ngers that differ significantly in one element also differ consistently in others within the same nucleosynthetic family, then one could conclude that this nucleosynthetic channel is driving the variations. 

\begin{figure*}
    \centering
    \includegraphics[width=\linewidth]{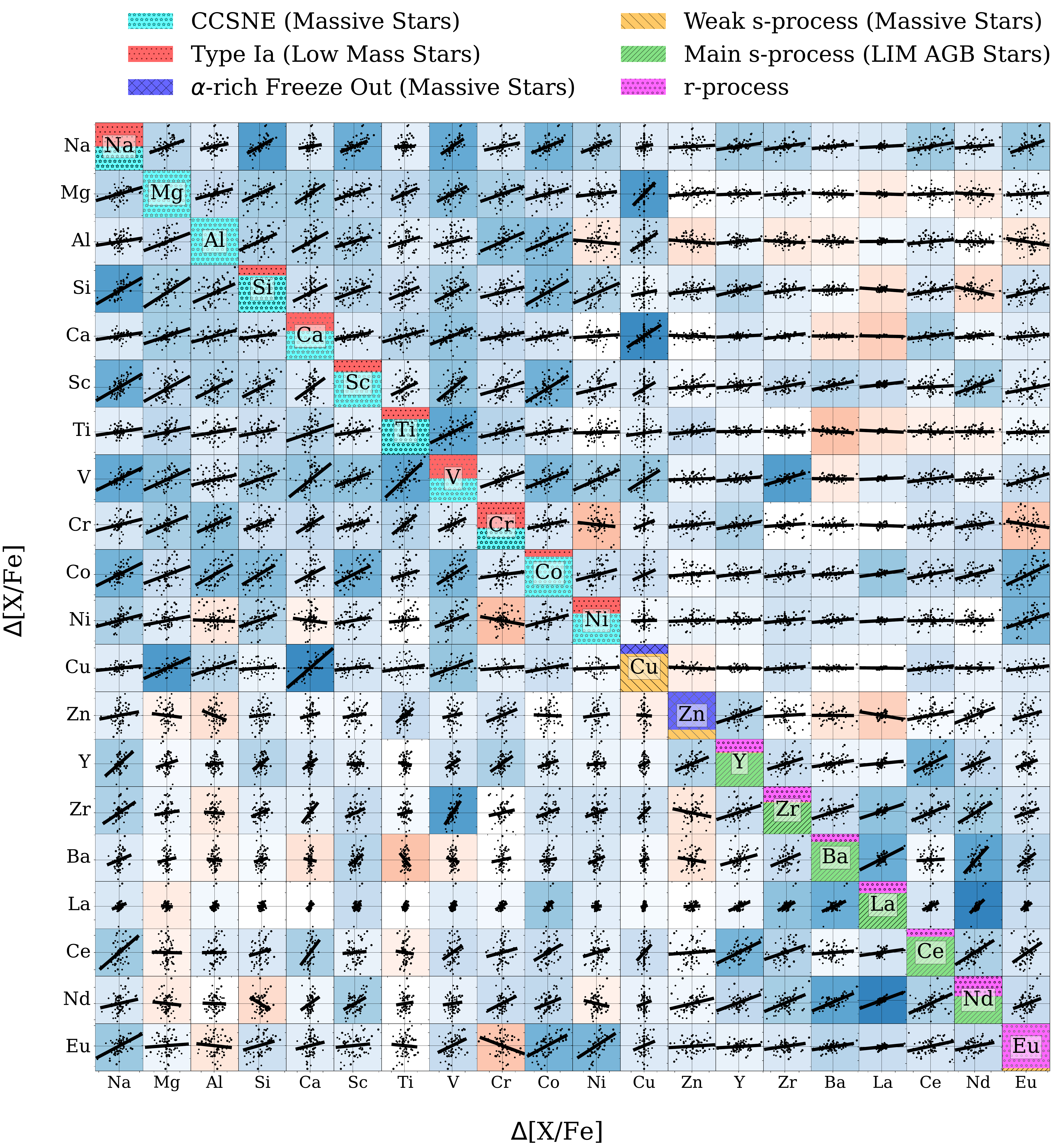}
    \caption{A comparison of doppelg{\"a}ngers' abundance difference ($\Delta$[X/Fe]) in all combinations of elements measured in the optical spectra.  In each panel, each pairs' abundance difference is reflected (shown twice) to illustrate the arbitrary ordering of the two stars. Lines represent the linear regression\footnote{We use the Huber Regressor from \code{scikit-learn} which is less influenced by outliers than other methods.} through the data in each gridbox, colors indicate the sign of the slope (blue for positive, red for negative), and transparency indicates the magnitude of the Pearson correlation coefficient.  Diagonal elements are colored by relative contribution from various nucleosynthetic sources to the Sun's abundance in each element \citep{Bisterzo2014, Kobayashi2020}.  Of the newly-measured elements (Cu through Eu), heavy \textit{s}-process elements Ba, La, Ce, and Nd show the strongest correlations with each other, indicating that stellar pairs (dis)similar in one of these elements tend to be (dis)similar in the others.}
    \label{fig:ncap_cov}
\end{figure*}

Figure \ref{fig:ncap_cov} compares doppelg{\"a}ngers' similarities in one element to those in other elements.  Note that all points are reflected to illustrate the arbitrary ordering of stars in each pair. We find positive correlations between pairs' similarities in most light-, $\alpha$, odd-Z, and Fe-peak elements. Though these abundance differences are small, correlations are still detected.  Moving onto the newly-measured elements, significant correlations exist between Cu, Mg, and Ca.  We find strong correlations between heavy \textit{s}-process elements Ba, La, Ce, and Nd, four elements that we found earlier can differ among doppelg{\"a}ngers (though Ce notably shows the weakest correlations with the other elements of the heavy s-process group).  We find comparatively weaker correlations between light \textit{s}- elements Y and Zr and the rest of the \textit{s}-process elements with the exception of Y and Ce.  Correlations between light (Z \textless 29) and neutron-capture elements also exist.  We see correlations between Zr and V and also weak ones between Eu, Co, and Ni.

\begin{figure}
    \centering
    \includegraphics[width=\linewidth]{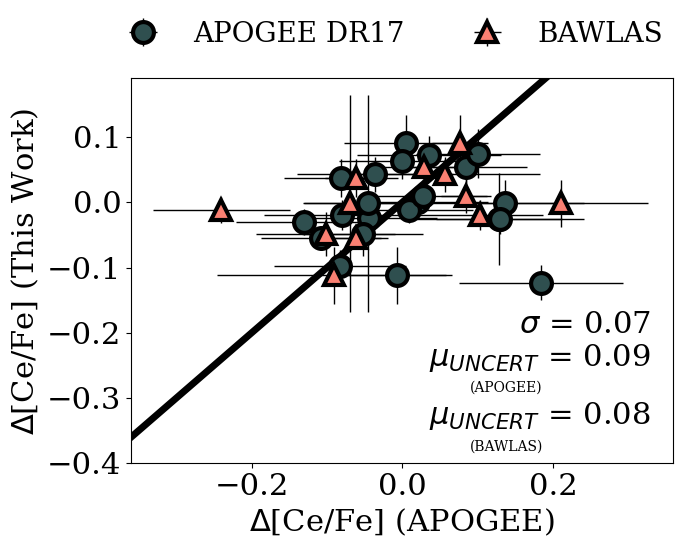}
    \caption{A comparison of this work's $\Delta$ [Ce/Fe] and those reported by \textsl{APOGEE} DR17 (navy circles) and \textsl{BAWLAS} (red triangles) for chemical doppelg{\"a}ngers in this sample.  The one-to-one line is included in solid black.  Our results generally agree within measurement uncertainties barring two exceptions.  The Ce abundance uncertainties reported by \textsl{APOGEE} and \textsl{BAWLAS} exceed the median abundance differences between chemical doppelg{\"a}ngers measured in our optical analysis, hence why these stars were reported as doppelg{\"a}ngers in our initial search.  However, our higher precision optical analsis is able to distinguish between some doppelg{\"a}ngers using [Ce/Fe] and other neutron-capture elements.}
    \label{fig:ce}
\end{figure}

\subsection{[Ce/Fe] in \textsl{APOGEE} and \textsl{BAWLAS}}
Our analysis suggests that highly chemically similar stars in \textsl{APOGEE} can show significant differences in the optically-measured neutron-capture elements despite using \textsl{APOGEE}-reported Ce in our initial selection.  This suggests that \textsl{APOGEE}-measured Ce may not be sufficient for distinguishing between stars that are otherwise chemically similar.  To understand why this may be the case, we compare our optically-measured pairwise $\Delta$[Ce/Fe] to that reported by \textsl{APOGEE} and \textsl{BAWLAS} in Figure \ref{fig:ce}.  Our measurements of $\Delta$[Ce/Fe] generally agree with \textsl{APOGEE}'s and \textsl{BAWLAS}'s within uncertainties with a few exceptions.  However, \textsl{APOGEE}'s and \textsl{BAWLAS}'s precisions in [Ce/Fe], which are on average 0.09 dex and 0.08 dex, respectively, exceeds the typical neutron-capture element differences of our sample (\textless 0.02-0.05 dex).  Thus, compared to optically-measured neutron-capture elements, \textsl{APOGEE}'s and \textsl{BAWLAS}'s [Ce/Fe] measurements are not sufficiently precise for resolving abundance differences that can exist among stars that are otherwise chemically indistinguishable in the lighter elements.

\section{Discussion}\label{sec:discussion}

\subsection{Summary and Comparison to Prior Works}
This work shows that \textsl{APOGEE}-DR17-identified "chemical doppelg{\"a}ngers" (see definition in Section \ref{sec:methods}) are not necessarily doppelg{\"a}ngers in the neutron-capture elements when studied in the optical at high resolution.  In other words, there exist highly chemically similar disk stars that share nearly indistinguishable lighter (Z\textless29) element abundances, nearly identical C/N ratios (a proxy for age in giants with similar stellar parameters, e.g., \citealt{Spoo2022}) but differ measurably in these elements.  We emphasize that these differences are on average small ($\sim$ 0.02 to 0.05 dex beyond the typical scatter in M~67) but can reach over 0.20 and up to 0.38 dex.  All 25 pairs have differences at this level in at least one neutron-capture element. This demonstrates that \textsl{APOGEE} DR17 abundances, even when measured for 17 elements from SNR \textgreater 300 spectra, are not always sufficient for identifying truly chemically similar stars in 8 additional elements. That is, we find that the neutron-capture elements contribute additional information not captured by the lighter elements.

From a Galactic chemical evolution (GCE) point of view, these results demonstrate that the neutron-capture elements operate semi-independently from the lighter elements insofar as a giant star's lighter element abundances (Z\textless29), including C and N, do not perfectly dictate its heavy element abundances.  This semi-independence of the GCE of neutron-capture elements is supported by observations. For example, \citealt{Bedell2018} performed a high precision differential abundance analysis of Solar twins in the Solar neighborhood.  Figure 3 of their work shows that at fixed stellar age, Sun-like stars in the Solar neighborhood have a larger scatter in the abundances of Cu and neutron-capture elements than lighter elements. Griffith \& Blum, in prep., perform a similar study in the low metallicity regime and also find that \textsl{s}-process elements show larger star-to-star scatter than the lighter elements. This suggests that in both the low and Solar metallicity regimes, stars born at the same time can display a wider variety of Cu and neutron-capture element compositions than light element compositions.   Furthermore, \citet[][]{Griffith2022, Griffith2024} show that a two-process Galactic chemical evolution model described by relative contributions from Type-Ia and Type-II supernovae is not sufficient to predict the abundances of elements produced via the neutron-capture process.  \citet{Mead2025} similarly finds that the lighter elements can not predict the neutron-capture elemental abundances of Solar neighborhood stars beyond a precision of 0.02 to 0.05 dex, even when using abundances derived from high SNR (SNR \textgreater 1000), high resolution (R$\sim$110,000) spectra. All of these results support the semi-independent GCE trajectories of these heavier elements. However, these amplitudes od 0.02-0.05 dex also represent the precision required on the measurements to access this independent information which is otherwise obfuscated by noise.

\subsection{Distinguishing power of individual elements and nucleosynthetic channels}
The distinct nucleosynthetic processes and sites that create the neutron-capture elements could explain why these elements are not perfectly traced by the lighter ones.  The following subsections briefly review the nucleosynthetic origins of each element newly measured in this work and dissect our results on an element-by-element basis.

\vspace{.5cm}

\subsubsection{Weak \textsl{s}-process elements: Copper and Zinc}
Our optical analysis reveals that \textsl{APOGEE} DR17-identified chemical doppelg{\"a}ngers on average do not differ significantly in [Cu or Zn/Fe].  Their nucleosynthetic origins could provide an explanation.  Though these elements (atomic numbers 29 and 30) are commonly associated with the Fe-peak group, their primary production mechanisms have been debated for decades \citep[see discussion in][]{Caffau2023}.  Cu was originally believed to be produced primarily via explosive nucleosynthesis in Type Ia and Type II supernovae \citep{Matteucci1993}.  \citet{Sneden1991} were among the first to propose that the weak \textit{s}-process in massive stars is the main source for the Sun's Cu abundance, and this was re-proposed by \citet{Bisterzo2005}.  Updates to neutron-capture reaction rates confirmed that in the metal-rich regime, Cu is primarily synthesized in massive stars (M \textgreater 8 $\rm M_\odot$) via the weak \textit{s}-process \citet{Heil2008}, and \citet{Pignatari2010} suggest that 70-80\% of Cu in the Sun was likely synthesized via this process.  The weak \textit{s}-process in massive stars operates during both the He core and C shell burning stages and produces elements between Fe and Mo (including Cu, Zn, Sr, Zr, and Y) in the periodic table.  The free neutrons required for the weak \textit{s}-process come from the $\rm ^{22}N(\alpha, n)^{25}Mg$ reaction which takes place primarily during the star's core He and C-burning phases \citep[e.g.,][]{Pignatari2010, Frischknect2016, Limongi2018}.  Additionally, \citet[e.g.,][]{Karakas2016} find that Cu can also be synthesized in $\sim$ Solar metallicity M~\textgreater~3$\rm M_{\odot}$ AGB stars via the \textit{s}-process, though it is believed that only 5\% of the Sun's Cu was produced in AGB stars \citep{Travaglio2004, Bisterzo2005}. Cu's primary dispersal mechanisms involve supernovae of various flavors, though it also enters the ISM via AGB winds.  The element's primarily massive star origin leads it to show a strong [X/Fe] trend with stellar age in the Solar neighborhood that mimics those of the alpha elements \citep[e.g.,][]{Spina2016, Bedell2018}.  

The production site of Zn is similarly complicated \citep[see][and references therein]{Hirai2018}.  As for Cu, the weak \textit{s}-process in massive stars produces Zn \citep[e.g.,][]{Woolsey1995}.  However, Zn is also produced in large quantities during alpha-rich freeze out, a process that occurs during the core-collapse explosion of a massive star.  The Si-rich shell of the massive star is struck by the supernova shock, and its contents break down into alpha particles and other nucleons. When this material cools, the free alpha particles and nucleons reassemble into heavier elements including Zn \citep[e.g.,][]{Woolsey1995, Jordan2003}.  Finally, hypernovae (explosions with $\sim$ 1 dex greater kinetic energy than typical core-collapse supernovae) and electron-capture supernovae are also responsible for the Galactic reservoir of Zn \citep[e.g.,][]{Kobayashi2006}.  In the Sun, alpha-rich freeze out is likely responsible for $\sim$ 80\% of its Zn content while the weak \textit{s}-process is responsible for the remaining 20\% \citep{Bisterzo2014}.  Unlike in the case of Cu, Zn does not appear to be produced in AGB stars \citep{Karakas2016}.

The unique nucleosynthetic origins of Cu and Zn, distinct from those of the light, $\alpha$, and Fe-peak elements, make them potentially important elements for distinguishing among doppelg{\"a}ngers.  However, evidently, doppelg{\"a}ngers on average do not differ in these elements.  In the context of this work, elements produced primarily via the weak \textit{s}-process in massive stars have no additional distinguishing power beyond the lighter elements.  The shared dispersal mechanisms of Cu, Zn, and the lighter elements, namely supernovae, could be responsible (see discussion in Section \ref{subsec:theory2} regarding implications for the mixing length of supernova products).

\vspace{.5cm}

\subsubsection{Light \textsl{s}-process elements: Yttrium and Zirconium}

Our results suggest that \textsl{APOGEE}-identified doppelg{\"a}ngers can differ significantly in [Y and Zr/Fe]. These elements are associated with the first \textit{s}-process peak and referred to as light \textit{s}-process (light \textit{s}-) elements.  Light \textit{s}- elements are synthesized via both the weak \textit{s}-process in massive stars (see discussion in previous subsection) and the main \textit{s}-process in low and intermediate mass (LIM, M~\textless~8 M$_\odot$) AGB stars, where the required source of free neutrons comes from the $\rm ^{13}C(\alpha, n)^{16}O$ reaction within the He-intershell \citep[e.g.,][]{Cristallo2011, Longland2012, Karakas2016, Cristallo2018}.  Products of the neutron-capture process are then dredged up to the surface before being expelled via its stellar wind.  In AGB stars, neutron-capture nucleosynthesis and dredge up are part of a cyclical process that occurs during the thermally pulsing AGB phase.  AGB stars can experience tens to hundreds of thermal pulses before neutron-capture nucleosynthesis ceases \citep[e.g.,][]{Karakas2014}.  As with all neutron-capture elements, Y and Zr are not exclusively formed via the \textit{s}-process and can also be synthesized via the \textit{r}-process.  Recent works estimate that 66\% of Zr and 72\% of Y in the Solar system were formed via the \textit{s}-process, respectively \citep{Bisterzo2014, Kobayashi2020}, with the rest synthesized via the \textit{r}-process (see Section \ref{sec:Eu} discussing Eu).  Despite belonging to the same nucleosynthetic group, Y and Zr behave differently in the Galactic disk.  Though \citet{Bedell2018} show that the [X/Fe] vs. age trends for Y and Zr appear indistinguishable, Fig. 11 in \citet{Delgado2017} illustrates that [Zr/Fe] shows a strong inverse correlation with [Fe/H] while [Y/Fe] remains relatively constant.  This could point to a greater contribution to Galactic [Zr/Fe] from massive stars.

Zr and Y differ on average by 0.035 dex, 0.02 to 0.03 dex greater than M~67 pairs.  As mentioned above, these elements have significant contributions from the main \textit{s}-process in AGB stars, suggesting that this nucleosynthetic source could increase the chemical inhomogeneity of the disk.

\vspace{.5cm}

\subsubsection{Heavy \textsl{s}-process elements: Lanthanum and Cerium}

Ba, La, Ce vary most significantly among \textsl{APOGEE}-identified chemical doppelg{\"a}ngers, on average by 0.03 to 0.048 dex and by over 0.25 dex for Pair 47. These elements are considered among the purest \textit{s}-process elements with the lowest contributions from \textit{r}-process nucleosynthetic sources.  Theoretical and observational works suggest that 85\% of Ba, 76\% of La and 84\% of Ce in the Solar System was produced via the \textit{s}-process, with the main \textit{s}-process in low and intermediate mass AGB stars (see previous subsection) dominating production \citep[e.g.,][]{Travaglio1999, Winckler2006, Serminato2009, Bisterzo2014, Karakas2014, Kobayashi2020}.  [Ba, La and Ce/Fe] show the steepest gradients with stellar age in the Solar neighborhood likely due to their low mass AGB star origins \citep{Bedell2018}.  The time delay between the production of Fe, which has significant contributions from high mass, short lived stars that end as Type II supernovae and that of long-lived AGB stars causes steep abundance gradients with stellar age.  This principle explains why heavy \textit{s}- elements, when compared to pure Type-II elements like Mg, are the strongest chemical clocks \citep[e.g.,][]{Ratcliffe2024}.  Ba, La, Ce, and Y variations among doppelg{\"a}ngers indicate that AGB star nucleosynthesis likely plays a role in abundance variations among otherwise chemically indistinguishable stars.


\vspace{.5cm}

\subsubsection{Mixed element: Neodymium}

Our results show that Nd can sometimes distinguish between \textsl{APOGEE}-identified doppelg{\"a}ngers, on average showing differences 0.02 dex greater than those among M~67 pairs.  Nd is considered a mixed element as it has significant contributions from both \textit{s}- and \textit{r}-process nucleosynthesis.  In our Solar System, 56\% of Nd was produced through the \textit{s}-process in low-mass AGB stars while the rest was created through the \textit{r}-process \citep{Bisterzo2014, Kobayashi2020}.  See the next subsection for a discussion of the \textit{r}-process and its astrophysical sources. Given our results for Y, Zr, Ba, La, and Ce, it is possible that Nd's AGB star contributions could be responsible for its enhanced distinguishing power.

\vspace{.5cm}

\subsubsection{\textsl{r}-process element: Europium} \label{sec:Eu}
This work finds that \textsl{APOGEE}-identified chemical doppelg{\"a}ngers can differ in [Eu/Fe] beyond abundance uncertainties 20\% of the time.  However, these differences are generally small, on average exceeding the abundance differences between pairs of M~67 stars by just 0.01 dex.

Eu is the most accessible \textit{r}-process element in Solar metallicity stellar spectra \citep[e.g.,][]{GALAHDR4}.  The \textit{r}-process is responsible for 94\% of Solar Eu \citep{Bisterzo2014, Kobayashi2020}, and there is significant debate surrounding its astrophysical sites. GCE trends for Eu suggest that \textit{r}-process nucleosynthesis must be associated with both massive star evolution and neutron star mergers \citep[e.g.,][]{Matteucci2014, Delgado2017}. Recently, observations of a gravitational wave event generated by a neutron star merger identified \textit{r}-process elements in the optical spectrum, confirming the production of these elements through this channel \citep{Cote2018}.  \textit{r}-process elements have the greatest potential for stochastic variations among otherwise chemically similar field stars due to their nucleosynthetic sources being highly productive, energetic, and rare.  Among our disk star sample, it appears that \textit{r}-process products on average do not differ as significantly as \textit{s}-process elements among otherwise chemically similar stars.  However, they can still differentiate between one in five doppelgangers.  Importantly, we find two pairs where [Eu/Fe] differs by up to 0.10 dex despite comparatively weaker \textit{s}-process element differences.  This result suggests that Eu can be considered as an additional, although typically less effective, dimension when identifying chemically (dis)similar disk stars, supporting previous work identifying [Eu/$\alpha$] as an effective chemical tag \citep[e.g.,][]{Monty2024}.

Our results suggest that products of AGB star nucleosynthesis are most important for distinguishing among otherwise chemically similar disk stars.  By contrast, Cu, Zn, and Eu, all elements that enter the ISM via supernovae, do not share this distinguishing power, although [Eu/Fe] can occasionally distinguish among pairs, something that could be attributed to the stochasticity of its production.  The following two subsections present possible physical explanations for this phenomenon.

\subsection{Possible Physical  Interpretations}

\subsubsection{Possibility 1: High resolution, Optically-measured Neutron-Capture Elements May Provide Better Constraints on Stellar Age than \textsl{APOGEE} Abundances Alone}\label{subsec:theory1}

One possible explanation for why some \textsl{APOGEE}-identified doppelg{\"a}ngers differ in their \textit{s}-process neutron-capture abundances despite sharing indistinguishable lighter element abundances could be rooted in the relationship between \textit{s}-process elements and age.  \textit{s}-process elemental abundance ratios ([s/$\alpha$] and [s/Al]) have been found to correlate strongly with stellar age, particularly in the Solar neighborhood and outer Milky Way disk \citep[e.g.,][]{Casali2020, Carbajo2024, Ratcliffe2024, Casali2025}.  It is therefore possible that doppelg{\"a}ngers with neutron-capture elemental abundance differences also differ in age.  


Under this assumption, our results may demonstrate that \textit{s}-process elemental abundances are better at identifying stars with similar ages than \textsl{APOGEE} DR17 abundances alone, even those measured from SNR \textgreater 300 spectra. These results could indicate that [C/N] and [$\alpha$/Fe], two chemical clocks available through \textsl{APOGEE}, are not as effective as [s/$\alpha$ or Al] at identifying coeval stars.  While it has been shown that [$\alpha$/Fe] is a weaker chemical clock \citep[e.g.,][]{Carbajo2024} than [s/$\alpha$], this has not been shown for [C/N] to our knowledge.  The age-[C/N] relationship measured in \citealt{Spoo2022}, which was constrained using open clusters in \textsl{APOGEE} DR17, shows a comparable slope and scatter to that of the age-[s/$\alpha$] trend measured in, e.g., \citealt{Ratcliffe2024}, which constrained the relationship using disk stars observed by GALAH.  It has been shown, however, that the [C/N] chemical clock is unreliable in stars that experience extra mixing \citep[e.g.,][]{Shetrone2019}.  It is possible that this extra mixing could affect the [C/N] ratios of some doppelg{\"a}ngers in our sample, and \textit{s}-process elements, which do not suffer from this effect, are more reliable at tracing age in these cases.

If this explanation holds true, then it is possible that combining \textsl{APOGEE} abundances with GALAH- or Gaia-ESO-measured \textit{s}-process elements (Zr, Y, Ba, La, Ce, and Nd) could better identify stars born at similar Galactic times than \textsl{APOGEE} abundances alone.  These results suggest that when working with \textsl{APOGEE} abundances derived from SNR \textgreater 300 spectra, \textit{s}-process elemental abundance precisions of $\sim$ 0.02 to 0.050 dex would be most beneficial for resolving typical neutron-capture abundance differences among \textsl{APOGEE} disk stars.  Achieving such high precision is difficult without line-by-line differential analysis.  Lower precision abundances will still be helpful, though, as some doppelg{\"a}ngers in our sample differ by over 0.2 dex in \textit{s}-process elements.

\subsubsection{Possibility 2: High resolution, Optically-measured Neutron-Capture Elements May Provide Better Constraints on Galactic Birth Location than \textsl{APOGEE} Abundances Alone} \label{subsec:theory2}
While differences in stellar age could explain doppelg{\"a}ngers' differences in neutron-capture elements, it is also possible that age is not a factor.  If we assume that \textsl{APOGEE} C/N is as precise an age indicator as [s/$\alpha$ or Al], as suggested by previous works \citep[e.g.,][]{Casali2019, Spoo2022}, and that stellar birth radius is traced by stellar age and metallicity (\citealp{Ness2019, Lu2022}), then we can assume that the \textsl{APOGEE}-identified doppelg{\"a}ngers likely formed at the same time and Galactocentric birth radius.  By studying the Cu, Zn, and neutron-capture elemental abundance differences among APOGEE doppelg{\"a}ngers, we are effectively exploring the behavior of these elements in roughly mono-age, mono-birth radius stellar populations.  Under these assumptions, our results could indicate that neutron-capture elements show additional abundance variations in volumes of the ISM that are otherwise chemically homogeneous.  This could indicate that similarity in neutron-capture elemental abundances may trace stars' similarity in Galactic birth location more finely than the lighter elements alone.  


Adopting this framework, doppelg{\"a}ngers with similar \textit{s}-process elemental abundances may have formed at more similar Galactocentric radii or more similar Galactocentric azimuths than doppelg{\"a}ngers that differ in their \textit{s}-process elemental compositions.  If azimuthal mixing of AGB star products is efficient, as suggested by some simulations \citep[e.g.][]{Zhang2025}, then the former scenario is more likely.  In this scenario, neutron-capture elements are well-mixed azimuthally but show greater variations radially (i.e., steeper radial gradients) in the ISM relative to the elements dispersed by supernovae. Evidence supporting this can be found in, e.g., \citet{Molero2023}, which infers Galactic gradients of [La/H] and [Ce/H] up to 0.02 dex kpc$^{-1}$ steeper than those of [Fe/H].  

\subsection{Optical Spectroscopy Complements H-band Spectroscopy}
Our optical analysis suggests that \textsl{APOGEE} abundances do not tell a star's full nucleosynthetic story, even when SNR \textgreater 300 spectra and [C/N] are available.  Stars that \textsl{APOGEE} reports to be chemically indistinguishable can differ significantly in the elements with large \textit{s}-process contributions (e.g., Y, Ba, La, Ce, and Nd). Evidently, this is true even when [Ce/Fe] measurements from the SNR \textgreater 300 \textsl{APOGEE} spectra are available. At this SNR, [Ce/Fe] can be measured with an average precision of 0.08 dex from \textsl{APOGEE} spectra (Figure \ref{fig:ce}).  However, according to our results, this exceeds the typical star-to-star Ce variations that \textsl{APOGEE} doppelg{\"a}ngers can display, so one cannot use \textsl{APOGEE} [Ce/Fe] to further distinguish between disk stars beyond what the lighter (Z \textless 29) elements already accomplish.  This is further confirmed by testing the impact of including vs. excluding [Ce/Fe] when measuring the \textsl{APOGEE} DR17 doppelg{\"a}nger rate (a test we performed when creating Figure \ref{fig:apogee_dr}).  There is no difference in the measured doppelg{\"a}nger rate when including Ce.

Achieving the necessary neutron-capture element precision to further distinguish between stars is comparatively easier in the optical spectral regime where strong lines of the neutron-capture elements exist.  For example, at comparable resolving power to \textsl{APOGEE} but one third the SNR, one can measure [Ce/Fe] precisions between 0.03 and 0.05 dex in optical spectra \citep[e.g.,][]{GALAHDR3, gaiaeso}, enough to distinguish between \textsl{APOGEE}-identified chemically similar stars.  Furthermore, in the optical regime, several elements of the \textit{s}-process family beyond Ce can be measured in tandem to further denoise the \textit{s}-process signal of each star.  

This work demonstrates that high resolution, high signal-to-noise optical spectroscopy of a small sample of stars can reveal important information about stellar nucleosynthesis and Galactic chemical evolution.  Although surveys such as \textsl{APOGEE/SDSS-V} are observing millions of stars, much can be learned from a sample of just a few dozen with high fidelity abundances \citep[e.g.,][]{Spina2018, Bedell2018, Mead2025}.

\section{Conclusions}\label{sec:conclusion}

In this work, we assess the chemical similarity of \textsl{APOGEE}-identified "chemical doppelg{\"a}ngers" though the lens of optical, high resolution spectroscopy.  This point of view allows us to confirm their chemical similarity in the \textsl{APOGEE}-measured elements while also studying their abundances of Cu, Zn, and neutron-capture elements that are imprecisely (if at all) measured by the infrared survey.  Our line-by-line differential analysis reveals that stars deemed by \textsl{APOGEE} DR17 to be chemical doppelg{\"a}ngers according to their light (C, N, O), $\alpha$ (Mg, Si, Ca), Fe-peak (Fe, Ni, Co, V, Sc, Cr), and odd-Z (Na, Al, S, K, Mn) elements can differ detectably in neutron-capture elements Zr, Y, Ba, La, Ce, Nd, and occasionally Eu.  These differences range from $\Delta$[X/Fe] = 0.020$\pm$0.015 to 0.380$\pm$0.15 dex (4-140\%), and up to 0.05 dex (12\%) on average.  We find that doppelg{\"a}ngers that differ in one \textit{s}-process element tend to differ in other \textit{s}-process elements, though not always in equal magnitude.  Additionally, we show that Ce measured from \textsl{APOGEE} spectra is not sufficient for resolving differences between otherwise chemically indistinguishable stars, but neutron-capture elements measured from optical, high resolution spectra can resolve these differences. This work indicates that neutron-capture elements can aid in searches for chemically and (potentially) dynamically similar stars when used in tandem with the lighter (Z \textless 29) elements.  Furthermore, our results suggest that even at fixed [C/N] and light element composition, stars can differ in neutron-capture elemental abundances.  We discuss possible interpretations which include imperfections with the [C/N] chemical clock that can be addressed with \textit{s}-process-based chemical clocks and comparatively less efficient radial mixing of AGB star nucleosynthetic products.


\section*{Acknowledgements}

The authors thank the reviewer for their careful review of this manuscript and helpful comments that improved the quality of this work.  CM thanks Harriet Dinerstein, Adrian Price-Whelan, Adam Wheeler, Natalie Myers, Amaya Sinha, Laia Casamiquela, and David Weinberg for fruitful discussions that enhanced the determination and interpretation of these results. CM thanks the Center for Computational Astrophysics' Astro Data, Nearby Universe, and Spectroscopy groups for faciliating group discussions that enhanced the interpretation of this work.  CM thanks McDonald Observatory Staff Coyne Gibson, Phillip Macqueen, John Kuehne, Karen Sulewski, Patricia Granado, and Marcela Garcia for enabling this science by guiding, teaching, feeding, and housing CM as she gathered the observational data for this work. CM is supported by the NSF Astronomy and Astrophysics Fellowship award number AST-2401638. KH is partially supported through the Wootton Center for Astrophysical Plasma Properties funded under the United States Department of Energy collaborative agreement DE-NA0003843. EJG is supported by an NSF Astronomy and Astrophysics Postdoctoral Fellowship under award AST-2202135. KVJ is supported by Simons Foundation grant 1018465.

The following software and programming languages made this research possible: Python (version 3.9) and its packages astropy (\citealt{astropy:2013, astropy:2018, astropy:2022}), scipy \citep{2020SciPy-NMeth}, matplotlib \citep{matplotlib}, pandas (version 0.20.2; \citealt{pandas}) and  NumPy \citep{numpy, 2020NumPy-Array}. This research has made use of the VizieR catalog access tool, CDS, Strasbourg, France. The original description of the VizieR service was published in A\&AS 143, 23.

This work has made use of data from the European Space Agency (ESA) mission \textit{Gaia} (\url{https://www.cosmos. esa.int/gaia}), processed by the \textit{Gaia} Data Processing and Analysis Consortium (DPAC, \url{https://www.cosmos.esa.int/web/gaia/dpac/consortium}). Funding for the DPAC has been provided by national institutions, in particular the institutions participating in the \textit{Gaia} Multilateral Agreement.


\bibliographystyle{mnras}
\bibliography{main.bib} 

\appendix \label{appendix}
In this work, we adopt \textsl{APOGEE}-reported $\rm T_{eff}$ and logg values because they are derived from SNR \textgreater 300 spectra and exceed the precision we can achieve using BACCHUS on our optical spectra.  However, it is of interest to check whether the APOGEE parameters satisfy Fe ionization-excitation balance, the method used by BACCHUS to determine these parameters.  As this work is entirely differential, we check for the satisfaction of differential spectroscopic equilibrium (see, e.g., \citealp{Melendez2012, Yong2013, Melendez2014, Liu2021, Yong2023, Liu2024}, and references in \citealp{Nissen2018}), which requires a null trend between line-by-line $\Delta$[Fe/H] vs. excitation potential and an agreement between the average $\Delta$[Fe I/H] and $\Delta$[Fe II/H] values within 1 $\sigma$.

\begin{figure}
    \centering
    \includegraphics[width=1\linewidth]{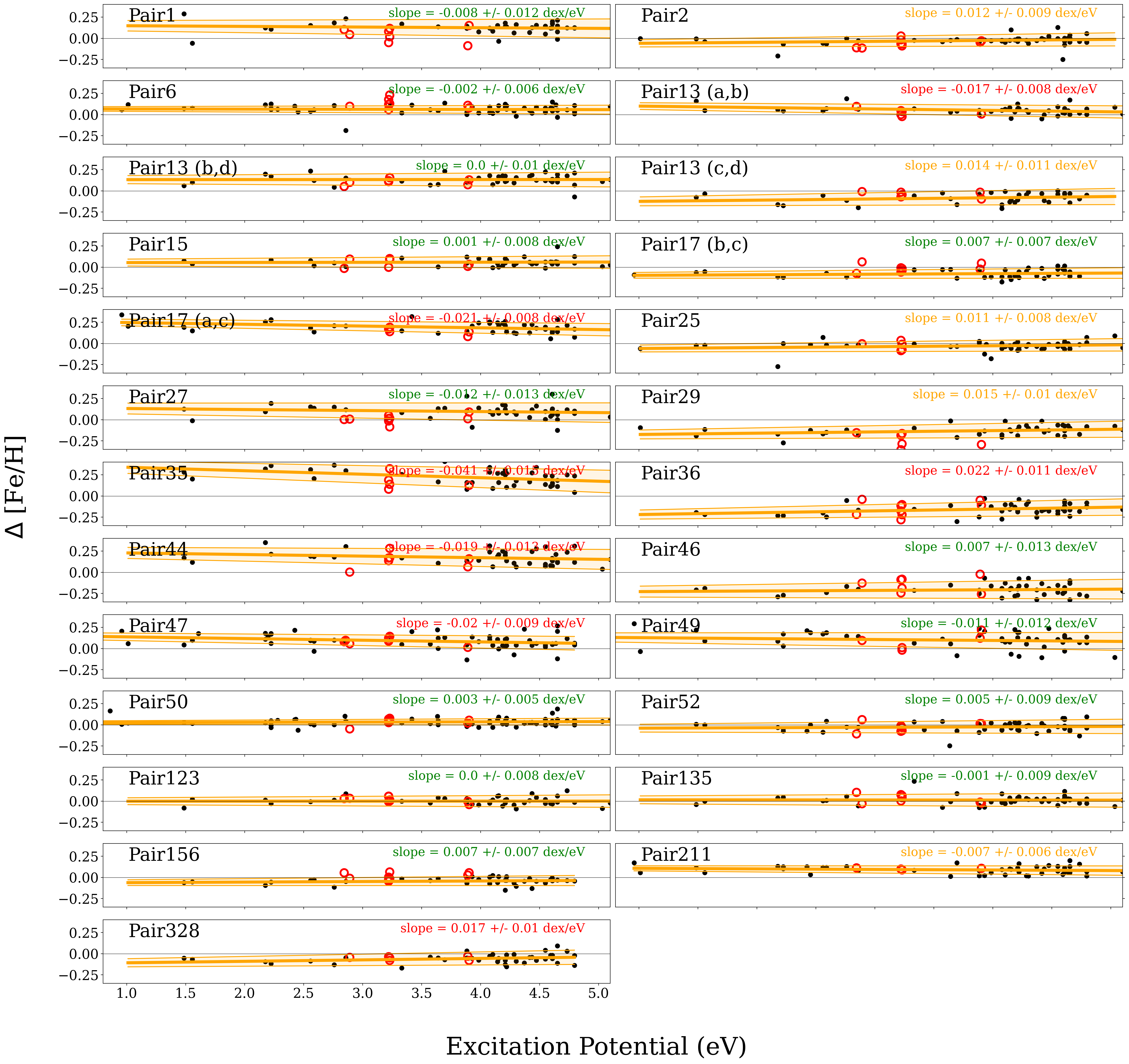}
    \caption{Line-by-line [Fe/H] differences among stars in each doppelganger pair as a function of excitation potential determined from Fe I (black points) Fe II (red circles) absorption lines.  The orange line and shaded region represents the best fit line through the black points, and the slope and associated uncertainty in the slope are printed within each panel. If \textsl{APOGEE}-reported $\rm T_{eff}$ and logg differences satisfy Fe ionization-excitation balance, then the slope of each line should be consistent with zero and average $\Delta$[Fe/H] abundances of the black and red markers should agree within 1$\sigma$.  When Fe excitation balance is satisfied (indicating suitable $\rm T_{eff}$), the text in the top right of each panel is colored green.  Otherwise, it is orange (would be satisfied with a \textless 25 K perturbation to $\rm T_{eff}$) or red (requires a larger perturbation of $\rm T_{eff}$).}
    \label{fig:ei}
\end{figure}

The results of our investigation are presented in Figure \ref{fig:ei}.  We find that in 13/25 pairs, Fe excitation balance is satisfied when adopting APOGEE’s $\rm T_{eff}$.  In an additional five pairs, Fe excitation balance would be satisfied within 0.005 dex/eV which (according to empirical tests reported in the BACCHUS manual) requires a \textless 25 K perturbation to $\rm T_{eff}$. In the remaining seven cases, APOGEE’s Teff does not satisfy excitation balance, and Teff would need to be shifted by \textgreater 25 K to satisfy it.  These pairs are Pairs 13ab, 17ac, 35, 36, 44, 47, and 328. The most extreme case is Pair 35, which would require an adjustment of $\sim$ 130 K to satisfy excitation balance. The rest of the pairs would require an adjustment of ~50 K.  As for logg, we find that APOGEE's logg satisfies ionization balance for all pairs.

This test indicates that APOGEE’s parameters do not always satisfy Fe excitation equilibrium, indicating that the line-by-line Fe ionization-excitation method with BACCHUS would yield slightly different $\rm T_{eff}$ values for a subset of our sample.  However, most importantly, we see no correlation between a lack of spectroscopic equilibrium satisfaction and the presence of significant s-process abundance differences, indicating that our results are not driven by the lack of excitation equilibrium.  With the exception of Zr, our neutron-capture element abundances are determined from singly ionized lines which are more sensitive to logg and ionization equilibrium.  The slight $\rm T_{eff}$ perturbations that would be needed to satisfy excitation balance in a subset of the pairs would not impact our results for these elements.   Furthermore, the ionization balance test has some limitations; it can have variable sensitivity and degeneracies between parameters as well as inherit systematics from 1D LTE model assumptions \citep[e.g.,][]{Jofre2019}.  As such, it is not clear that parameters determined using the Fe ionization-excitation method on optical spectra would exceed the quality of those determined from APOGEE SNR \textgreater 300 spectra.

\label{lastpage}
\end{document}